%%%%%%%%%%%%%%%%%%     PLAIN TEX FILE       %%%%%%%%%%%%%%%%%%
 %%%%%%%%%%%%%%%%%%  %%%%%%%%%%%%%%%%%%  %%%%%%%%%%%%%%%%%%  %%%%%%%%%%%%%%%%%%
 %%%%%%%%%%%%%%%%%%  %%%%%%%%%%%%%%%%%%  %%%%%%%%%%%%%%%%%%  %%%%%%%%%%%%%%%%%%
 %%%%%%%%%%%%%%%%%%  %%%%%%%%%%%%%%%%%%  %%%%%%%%%%%%%%%%%%  %%%%%%%%%%%%%%%%%%

 %%%%%%%%%%%%%%%%%%  tex macros for preprints, cm version %%%%%%%%%%%%%%
%                     (P. Ginsparg, last updated 9/91)
%                if confused, type `b' in response to query
%
%---------------------------------------------------------------------%
%% site dependent options:
%% \unredoffs and \redoffs define horizontal and vertical offsets
%% respectively for unreduced and reduced modes. \speclscape defines
%% the \special{} call that sets printer to landscape (sideways) mode.
%% from standard set below, leave uncommented as appropriate or redefine
%
%%% next 400dpi
%\def\unredoffs{} \def\redoffs{\voffset=-.31truein\hoffset=-.48truein}
%\def\speclscape{\special{landscape}}
%
%%% apple lw
\def\unredoffs{} \def\redoffs{\voffset=-.31truein\hoffset=-.59truein}
\def\speclscape{\special{ps: landscape}}
%
%%% qms lasergrafix:
%\def\unredoffs{} \def\redoffs{\voffset=-.4truein\hoffset=.125truein}
%\def\speclscape{\special{qms: landscape}}
%
%%% saclay A4 paper:
%\def\unredoffs{\hoffset-.14truein\voffset-.2truein}
%\def\redoffs{\voffset=-.55truein\hoffset=-.1truein} \def\speclscape{}
%
%---------------------------------------------------------------------%
%
\newbox\leftpage \newdimen\fullhsize \newdimen\hstitle \newdimen\hsbody
\tolerance=1000\hfuzz=2pt
\catcode`\@=11 % This allows us to modify PLAIN macros.
\def\bigans{b }
%\message{ big or little (b/l)? }\read-1 to\answ
\def\answ{b }
\ifx\answ\bigans\message{(This will come out unreduced.}
\magnification=1200\unredoffs\baselineskip=16pt plus 2pt minus 1pt
\hsbody=\hsize \hstitle=\hsize %take default values for unreduced format
\else\message{(This will be reduced.} \let\l@r=L
\magnification=1000\baselineskip=16pt plus 2pt minus 1pt \vsize=7truein
\redoffs \hstitle=8truein\hsbody=4.75truein\fullhsize=10truein\hsize=\hsbody
\output={\ifnum\pageno=0 %%% This is the HUTP version
  \shipout\vbox{\speclscape{\hsize\fullhsize\makeheadline}
    \hbox to \fullhsize{\hfill\pagebody\hfill}}\advancepageno
  \else
  \almostshipout{\leftline{\vbox{\pagebody\makefootline}}}\advancepageno
  \fi}
\def\almostshipout#1{\if L\l@r \count1=1 \message{[\the\count0.\the\count1]}
      \global\setbox\leftpage=#1 \global\let\l@r=R
 \else \count1=2
  \shipout\vbox{\speclscape{\hsize\fullhsize\makeheadline}
      \hbox to\fullhsize{\box\leftpage\hfil#1}}  \global\let\l@r=L\fi}
\fi
%---------------------------------------------------------------------
%
\newcount\yearltd\yearltd=\year\advance\yearltd by -1900

\def\Title#1#2{\nopagenumbers\abstractfont\hsize=\hstitle\rightline{#1}%
\vskip 1in\centerline{\titlefont #2}\abstractfont\vskip .5in\pageno=0}
\def\Date#1{\vfill\leftline{#1}\tenpoint\supereject\global\hsize=\hsbody%
\footline={\hss\tenrm\folio\hss}}%      restores pagenumbers
%
%       use following instead of \Date on the preliminary draft,
%       puts date/time on each page in big mode, writes labels in margins

\def\draftmode{\message{ DRAFTMODE }\def\draftdate{{\rm preliminary draft:
\number\month/\number\day/\number\yearltd\ \ \hourmin}}%
\headline={\hfil\draftdate}\writelabels\baselineskip=20pt plus 2pt minus 2pt
 {\count255=\time\divide\count255 by 60 \xdef\hourmin{\number\count255}
  \multiply\count255 by-60\advance\count255 by\time
  \xdef\hourmin{\hourmin:\ifnum\count255<10 0\fi\the\count255}}}
%       use \nolabels to get rid of eqn, ref, and fig labels in draft mode
\def\nolabels{\def\wrlabeL##1{}\def\eqlabeL##1{}\def\reflabeL##1{}}
\def\writelabels{\def\wrlabeL##1{\leavevmode\vadjust{\rlap{\smash%
{\line{{\escapechar=` \hfill\rlap{\sevenrm\hskip.03in\string##1}}}}}}}%
\def\eqlabeL##1{{\escapechar-1\rlap{\sevenrm\hskip.05in\string##1}}}%
\def\reflabeL##1{\noexpand\llap{\noexpand\sevenrm\string\string\string##1}}}
\nolabels
%
% tagged sec numbers
\global\newcount\secno \global\secno=0
\global\newcount\meqno \global\meqno=1
\def\newsec#1{\global\advance\secno by1\message{(\the\secno. #1)}
%\ifx\answ\bigans \vfill\eject \else \bigbreak\bigskip \fi  %if desired
\global\subsecno=0\eqnres@t\noindent{\bf\the\secno. #1}
\writetoca{{\secsym} {#1}}\par\nobreak\medskip\nobreak}
\def\eqnres@t{\xdef\secsym{\the\secno.}\global\meqno=1\bigbreak\bigskip}
\def\sequentialequations{\def\eqnres@t{\bigbreak}}\xdef\secsym{}
\global\newcount\subsecno \global\subsecno=0
\def\subsec#1{\global\advance\subsecno by1\message{(\secsym\the\subsecno.
#1)}
\ifnum\lastpenalty>9000\else\bigbreak\fi
\noindent{\it\secsym\the\subsecno. #1}\writetoca{\string\quad
{\secsym\the\subsecno.} {#1}}\par\nobreak\medskip\nobreak}
\def\appendix#1#2{\global\meqno=1\global\subsecno=0\xdef\secsym{\hbox{#1.}}
\bigbreak\bigskip\noindent{\bf Appendix #1. #2}\message{(#1. #2)}
\writetoca{Appendix {#1.} {#2}}\par\nobreak\medskip\nobreak}
%
%       \eqn\label{a+b=c}       gives displayed equation, numbered
%                               consecutively within sections.
%     \eqnn and \eqna define labels in advance (of eqalign?)
%
\def\eqnn#1{\xdef #1{(\secsym\the\meqno)}\writedef{#1\leftbracket#1}%
\global\advance\meqno by1\wrlabeL#1}
\def\eqna#1{\xdef #1##1{\hbox{$(\secsym\the\meqno##1)$}}
\writedef{#1\numbersign1\leftbracket#1{\numbersign1}}%
\global\advance\meqno by1\wrlabeL{#1$\{\}$}}
\def\eqn#1#2{\xdef #1{(\secsym\the\meqno)}\writedef{#1\leftbracket#1}%
\global\advance\meqno by1$$#2\eqno#1\eqlabeL#1$$}
%
%                            footnotes
\newskip\footskip\footskip14pt plus 1pt minus 1pt %sets footnote baselineskip
\def\footnotefont{\ninepoint}\def\f@t#1{\footnotefont #1\@foot}
\def\f@@t{\baselineskip\footskip\bgroup\footnotefont\aftergroup\@foot\let\next}
\setbox\strutbox=\hbox{\vrule height9.5pt depth4.5pt width0pt}
\global\newcount\ftno \global\ftno=0
\def\foot{\global\advance\ftno by1\footnote{$^{\the\ftno}$}}
%
%say \footend to put footnotes at end
%will cause problems if \ref used inside \foot, instead use \nref before
\newwrite\ftfile
\def\footend{\def\foot{\global\advance\ftno by1\chardef\wfile=\ftfile
$^{\the\ftno}$\ifnum\ftno=1\immediate\openout\ftfile=foots.tmp\fi%
\immediate\write\ftfile{\noexpand\smallskip%
\noexpand\item{f\the\ftno:\ }\pctsign}\findarg}%
\def\footatend{\vfill\eject\immediate\closeout\ftfile{\parindent=20pt
\centerline{\bf Footnotes}\nobreak\bigskip\input foots.tmp }}}
\def\footatend{}
%
%     \ref\label{text}
% generates a number, assigns it to \label, generates an entry.
% To list the refs on a separate page,  \listrefs
%
\global\newcount\refno \global\refno=1
\newwrite\rfile
\def\ref{[\the\refno]\nref}
\def\nref#1{\xdef#1{[\the\refno]}\writedef{#1\leftbracket#1}%
\ifnum\refno=1\immediate\openout\rfile=refs.tmp\fi
\global\advance\refno by1\chardef\wfile=\rfile\immediate
\write\rfile{\noexpand\item{#1\ }\reflabeL{#1\hskip.31in}\pctsign}\findarg}
%        horrible hack to sidestep tex \write limitation
\def\findarg#1#{\begingroup\obeylines\newlinechar=`\^^M\pass@rg}
{\obeylines\gdef\pass@rg#1{\writ@line\relax #1^^M\hbox{}^^M}%
\gdef\writ@line#1^^M{\expandafter\toks0\expandafter{\striprel@x #1}%
\edef\next{\the\toks0}\ifx\next\em@rk\let\next=\endgroup\else\ifx\next\empty%
\else\immediate\write\wfile{\the\toks0}\fi\let\next=\writ@line\fi\next\relax}}
\def\striprel@x#1{} \def\em@rk{\hbox{}}
\def\lref{\begingroup\obeylines\lr@f}
\def\lr@f#1#2{\gdef#1{\ref#1{#2}}\endgroup\unskip}
\def\semi{;\hfil\break}
\def\addref#1{\immediate\write\rfile{\noexpand\item{}#1}} %now unnecessary
\def\startrefs#1{\immediate\openout\rfile=refs.tmp\refno=#1}
\def\xref{\expandafter\xr@f}\def\xr@f[#1]{#1}
\def\refs#1{\count255=1[\r@fs #1{\hbox{}}]}
\def\r@fs#1{\ifx\und@fined#1\message{reflabel \string#1 is undefined.}%
\nref#1{need to supply reference \string#1.}\fi%
\vphantom{\hphantom{#1}}\edef\next{#1}\ifx\next\em@rk\def\next{}%
\else\ifx\next#1\ifodd\count255\relax\xref#1\count255=0\fi%
\else#1\count255=1\fi\let\next=\r@fs\fi\next}
%

%
% this is ugly, but moore insists
\newwrite\ffile\global\newcount\figno \global\figno=1
\def\fig{fig.~\the\figno\nfig}
\def\nfig#1{\xdef#1{fig.~\the\figno}%
\writedef{#1\leftbracket fig.\noexpand~\the\figno}%
\ifnum\figno=1\immediate\openout\ffile=figs.tmp\fi\chardef\wfile=\ffile%
\immediate\write\ffile{\noexpand\medskip\noexpand\item{Fig.\ \the\figno. }
\reflabeL{#1\hskip.55in}\pctsign}\global\advance\figno by1\findarg}
\def\xfig{\expandafter\xf@g}\def\xf@g fig.\penalty\@M\ {}
\def\figs#1{figs.~\f@gs #1{\hbox{}}}
\def\f@gs#1{\edef\next{#1}\ifx\next\em@rk\def\next{}\else
\ifx\next#1\xfig #1\else#1\fi\let\next=\f@gs\fi\next}
\newwrite\lfile
{\escapechar-1\xdef\pctsign{\string\%}\xdef\leftbracket{\string\{}
\xdef\rightbracket{\string\}}\xdef\numbersign{\string\#}}

\def\writestop{\def\writestoppt{\immediate\write\lfile{\string\pageno%
\the\pageno\string\startrefs\leftbracket\the\refno\rightbracket%
\string\def\string\secsym\leftbracket\secsym\rightbracket%
\string\secno\the\secno\string\meqno\the\meqno}\immediate\closeout\lfile}}
\def\writestoppt{}\def\writedef#1{}
\def\seclab#1{\xdef #1{\the\secno}\writedef{#1\leftbracket#1}\wrlabeL{#1=#1}}
\def\subseclab#1{\xdef #1{\secsym\the\subsecno}%
\writedef{#1\leftbracket#1}\wrlabeL{#1=#1}}
\newwrite\tfile \def\writetoca#1{}
\def\leaderfill{\leaders\hbox to 1em{\hss.\hss}\hfill}
%        use this to write file with table of contents
\def\writetoc{\immediate\openout\tfile=toc.tmp
   \def\writetoca##1{{\edef\next{\write\tfile{\noindent ##1
   \string\leaderfill {\noexpand\number\pageno} \par}}\next}}}
%       and this lists table of contents on second pass

%
\catcode`\@=12 % at signs are no longer letters
%
%        Unpleasantness in calling in abstract and title fonts
\edef\tfontsize{\ifx\answ\bigans scaled\magstep3\else scaled\magstep4\fi}
\font\titlerm=cmr10 \tfontsize \font\titlerms=cmr7 \tfontsize
\font\titlermss=cmr5 \tfontsize \font\titlei=cmmi10 \tfontsize
\font\titleis=cmmi7 \tfontsize \font\titleiss=cmmi5 \tfontsize
\font\titlesy=cmsy10 \tfontsize \font\titlesys=cmsy7 \tfontsize
\font\titlesyss=cmsy5 \tfontsize \font\titleit=cmti10 \tfontsize
\skewchar\titlei='177 \skewchar\titleis='177 \skewchar\titleiss='177
\skewchar\titlesy='60 \skewchar\titlesys='60 \skewchar\titlesyss='60
\def\titlefont{\def\rm{\fam0\titlerm}% switch to title font
\textfont0=\titlerm \scriptfont0=\titlerms \scriptscriptfont0=\titlermss
\textfont1=\titlei \scriptfont1=\titleis \scriptscriptfont1=\titleiss
\textfont2=\titlesy \scriptfont2=\titlesys \scriptscriptfont2=\titlesyss
\textfont\itfam=\titleit \def\it{\fam\itfam\titleit}\rm}
 \ifx\answ\bigans\else scaled\magstep1\fi
\ifx\answ\bigans\def\abstractfont{\tenpoint}\else
\font\abssl=cmsl10 scaled \magstep1
\font\absrm=cmr10 scaled\magstep1 \font\absrms=cmr7 scaled\magstep1
\font\absrmss=cmr5 scaled\magstep1 \font\absi=cmmi10 scaled\magstep1
\font\absis=cmmi7 scaled\magstep1 \font\absiss=cmmi5 scaled\magstep1
\font\abssy=cmsy10 scaled\magstep1 \font\abssys=cmsy7 scaled\magstep1
\font\abssyss=cmsy5 scaled\magstep1 \font\absbf=cmbx10 scaled\magstep1
\skewchar\absi='177 \skewchar\absis='177 \skewchar\absiss='177
\skewchar\abssy='60 \skewchar\abssys='60 \skewchar\abssyss='60
\def\abstractfont{\def\rm{\fam0\absrm}% switch to abstract font
\textfont0=\absrm \scriptfont0=\absrms \scriptscriptfont0=\absrmss
\textfont1=\absi \scriptfont1=\absis \scriptscriptfont1=\absiss
\textfont2=\abssy \scriptfont2=\abssys \scriptscriptfont2=\abssyss
\textfont\itfam=\bigit \def\it{\fam\itfam\bigit}\def\footnotefont{\tenpoint}%
\textfont\slfam=\abssl \def\sl{\fam\slfam\abssl}%
\textfont\bffam=\absbf \def\bf{\fam\bffam\absbf}\rm}\fi
\def\tenpoint{\def\rm{\fam0\tenrm}% switch back to 10-point type
\textfont0=\tenrm \scriptfont0=\sevenrm \scriptscriptfont0=\fiverm
\textfont1=\teni  \scriptfont1=\seveni  \scriptscriptfont1=\fivei
\textfont2=\tensy \scriptfont2=\sevensy \scriptscriptfont2=\fivesy
\textfont\itfam=\tenit
\def\it{\fam\itfam\tenit}\def\footnotefont{\ninepoint}%
\textfont\bffam=\tenbf \def\bf{\fam\bffam\tenbf}\def\sl{\fam\slfam\tensl}\rm}
\font\ninerm=cmr9 \font\sixrm=cmr6 \font\ninei=cmmi9 \font\sixi=cmmi6
\font\ninesy=cmsy9 \font\sixsy=cmsy6 \font\ninebf=cmbx9
\font\nineit=cmti9 \font\ninesl=cmsl9 \skewchar\ninei='177
\skewchar\sixi='177 \skewchar\ninesy='60 \skewchar\sixsy='60
\def\ninepoint{\def\rm{\fam0\ninerm}% switch to footnote font
\textfont0=\ninerm \scriptfont0=\sixrm \scriptscriptfont0=\fiverm
\textfont1=\ninei \scriptfont1=\sixi \scriptscriptfont1=\fivei
\textfont2=\ninesy \scriptfont2=\sixsy \scriptscriptfont2=\fivesy
\textfont\itfam=\ninei \def\it{\fam\itfam\nineit}\def\sl{\fam\slfam\ninesl}%
\textfont\bffam=\ninebf \def\bf{\fam\bffam\ninebf}\rm}
%
%---------------------------------------------------------------------
%

\hyphenation{anom-aly anom-alies coun-ter-term coun-ter-terms}
\def\inv{^{\raise.15ex\hbox{${\scriptscriptstyle -}$}\kern-.05em 1}}

\def\Dsl{\,\raise.15ex\hbox{/}\mkern-13.5mu D} %this one can be subscripted
\def\dsl{\raise.15ex\hbox{/}\kern-.57em\partial}

 \def\Tr{{\rm Tr}}
\font\bigit=cmti10 scaled \magstep1
 %pound sterling
\def\lspace{\ifx\answ\bigans{}\else\qquad\fi}
\def\lbspace{\ifx\answ\bigans{}\else\hskip-.2in\fi} % $$\lbspace...$$
\def\boxeqn#1{\vcenter{\vbox{\hrule\hbox{\vrule\kern3pt\vbox{\kern3pt
           \hbox{${\displaystyle #1}$}\kern3pt}\kern3pt\vrule}\hrule}}}
\def\mbox#1#2{\vcenter{\hrule \hbox{\vrule height#2in
               \kern#1in \vrule} \hrule}}  %e.g. \mbox{.1}{.1}
%       matters of taste
%\def\tilde{\widetilde} \def\bar{\overline} \def\hat{\widehat}
%
% some sample definitions
  %     curly letters

\def\e#1{{\rm e}^{^{\textstyle#1}}}

\def\darr#1{\raise1.5ex\hbox{$\leftrightarrow$}\mkern-16.5mu #1}
 %pound sterling

\def\half{{\textstyle{1\over2}}} %puts a small half in a displayed eqn
\def\roughly#1{\raise.3ex\hbox{$#1$\kern-.75em\lower1ex\hbox{$\sim$}}}

%\input harvmac.tex

%%temporary additional macros
% \input macros.tex
% April 16 -- NN

%%%%%%%%%%%%%%%%%%%%%  Rublenye bukvy   %%%%%%%%%%%%%%%%%%%%%%%%
\def\IB{\relax\hbox{$\inbar\kern-.3em{\rm B}$}}
\def\IC{\relax\hbox{$\inbar\kern-.3em{\rm C}$}}
\def\ID{\relax\hbox{$\inbar\kern-.3em{\rm D}$}}
\def\IE{\relax\hbox{$\inbar\kern-.3em{\rm E}$}}
\def\IF{\relax\hbox{$\inbar\kern-.3em{\rm F}$}}
\def\IG{\relax\hbox{$\inbar\kern-.3em{\rm G}$}}
\def\IGa{\relax\hbox{${\rm I}\kern-.18em\Gamma$}}
\def\IH{\relax{\rm I\kern-.18em H}}
\def\IK{\relax{\rm I\kern-.18em K}}
\def\II{\relax{\rm I\kern-.18em I}}
\def\IL{\relax{\rm I\kern-.18em L}}
\def\IP{\relax{\rm I\kern-.18em P}}
\def\IR{\relax{\rm I\kern-.18em R}}
\def\IZ{\relax\ifmmode\mathchoice {\hbox{\cmss Z\kern-.4em Z}}{\hbox{\cmss
Z\kern-.4em Z}} {\lower.9pt\hbox{\cmsss Z\kern-.4em Z}}
{\lower1.2pt\hbox{\cmsss Z\kern-.4em Z}}\else{\cmss Z\kern-.4em Z}\fi}

\def\IB{\relax{\rm I\kern-.18em B}}
\def\IC{{\relax\hbox{$\inbar\kern-.3em{\rm C}$}}}
\def\ID{\relax{\rm I\kern-.18em D}}
\def\IE{\relax{\rm I\kern-.18em E}}
\def\IF{\relax{\rm I\kern-.18em F}}

%%%%%%%%%%%%%%%%%%%% Calligraphic letters  %%%%%%%%%%%%%%%%%%%%%%%

%%%%%%%%%%%%%%%%%%%%%%%%%% Derivatives  %%%%%%%%%%%%%%%%%%%%%%%%
\def\p{\partial}

%%Beltrami

%%%%%%%%%%%%%%%%%%%% letters with bar %%%%%%%%%%%%%%%%%%%%%%%%%%

%%%%%%%%%%%%%%%%%%%%%%%%%%% Math symbols %%%%%%%%%%%%%%%%%%%%%%%

\def\Tr{{\rm Tr}}

%%%%%%%%%%%%%%%%%%%%% Short Cuts %%%%%%%%%%%%%%%%%%%%%%%

\def\half {{1\over 2}}

%%%%%%%%%%%%%%%%%% Greek %%%%%%%%%%%%%%%%%%%%%%

\def\a{\alpha}
\def\b{\beta}
\def\g{\gamma}  
\def\d{\delta}  
\def\m{\mu}
\def\n{\nu}

\def\l{\lambda} \def\L{\Lambda}
\def\k{\kappa}
\def\e{\epsilon}

%%%%%%%%%%%%%%%%%% Big ( )  %%%%%%%%%%%%%%%%%%%%%%
\def\|{\Big|}
\def\({\Big(}   \def\){\Big)}
\def\[{\Big[}   \def\]{\Big]}

%%%%%%%%%%%%%%%%%% Text %%%%%%%%%%%%%%%%%%%%%%

%%%%%%%%%%%%% References %%%%%%%%%%%%%%%%%%%%

% refs with #1=authors, #2=title, #3=publ.ref, #4=hep no :
%\lref\NAME{\paper
%{Authors}{Title(in \it)}{\PLB{No.}{Year}{page},}
%{\hh 0006036 (in\tt)}.}

%\def\hh#1{hep-th/{\it #1}}

% journal~{\bf no.} (year) page

%%%%%%%%%%%%%%%%%%% Something to deal with sub-sub-sections
%%%%%%%%%%%%%%%%%%%%%%%%%%%%%%%%%%%%%%%%%%%%%%%

\def\unlockat{\catcode`\@=11}
\def\lockat{\catcode`\@=12}

\unlockat

% Something to deal with sub-sub-sections

\def\newsec#1{\global\advance\secno by1\message{(\the\secno. #1)}
\global\subsecno=0\global\subsubsecno=0\eqnres@t\noindent {\bf\the\secno. #1}
\writetoca{{\secsym} {#1}}\par\nobreak\medskip\nobreak}
\global\newcount\subsecno \global\subsecno=0
\def\subsec#1{\global\advance\subsecno by1\message{(\secsym\the\subsecno.
#1)}
\ifnum\lastpenalty>9000\else\bigbreak\fi\global\subsubsecno=0
\noindent{\it\secsym\the\subsecno. #1}
\writetoca{\string\quad {\secsym\the\subsecno.} {#1}}
\par\nobreak\medskip\nobreak}
\global\newcount\subsubsecno \global\subsubsecno=0
\def\subsubsec#1{\global\advance\subsubsecno by1
\message{(\secsym\the\subsecno.\the\subsubsecno. #1)}
\ifnum\lastpenalty>9000\else\bigbreak\fi
\noindent\quad{\secsym\the\subsecno.\the\subsubsecno.}{#1}
\writetoca{\string\qquad{\secsym\the\subsecno.\the\subsubsecno.}{#1}}
\par\nobreak\medskip\nobreak}

\def\subsubseclab#1{\DefWarn#1\xdef #1{\noexpand\hyperref{}{subsubsection}%
{\secsym\the\subsecno.\the\subsubsecno}%
{\secsym\the\subsecno.\the\subsubsecno}}%
\writedef{#1\leftbracket#1}\wrlabeL{#1=#1}}% Macros for boxes
\lockat

%why???\font\manual=manfnt
\def\dbend{\lower3.5pt\hbox{\manual\char127}}

%%%%%%%%%%%%%%%%%%% Macros for boxes %%%%%%%%%%%%%%%%%%

\def\boxit#1{\vbox{\hrule\hbox{\vrule\kern8pt
\vbox{\hbox{\kern8pt}\hbox{\vbox{#1}}\hbox{\kern8pt}}
\kern8pt\vrule}\hrule}}

\def\mathboxit#1{\vbox{\hrule\hbox{\vrule\kern8pt\vbox{\kern8pt
\hbox{$\displaystyle #1$}\kern8pt}\kern8pt\vrule}\hrule}}

\overfullrule=0pt

%%%%%%%%%%%%%%%%%%%%%%%%%% Derivatives  %%%%%%%%%%%%%%%%%%%%%%%%

\def\p{\partial}

%%Beltrami

%%%%%%%%%%%%%%%%%%%% letters with bar %%%%%%%%%%%%%%%%%%%%%%%%%%

%%%%%%%%%%%%%%%%%%%%% Short Cuts %%%%%%%%%%%%%%%%%%%%%%%

\def\half {{1\over 2}}

%%%%%%%%%%%%%%%%%% Greek %%%%%%%%%%%%%%%%%%%%%%

\def\a{\alpha}
\def\b{\beta}
\def\g{\gamma}  
\def\d{\delta}  
\def\m{\mu}
\def\n{\nu}

\def\l{\lambda} \def\L{\Lambda}
\def\k{\kappa}
\def\e{\epsilon}

\def\a{\alpha}
\def\b{\beta}
\def\d{\delta}

\def\m{\mu}
\def\n{\nu}

\def\l{\lambda}
\def\L{\Lambda}

\def\k{\kappa}

\def\t{\theta}

%%%%%%%%%%%%%%%%%% Big ( )  %%%%%%%%%%%%%%%%%%%%%%

\def\|{\Big|}
\def\({\Big(}   \def\){\Big)}
\def\[{\Big[}   \def\]{\Big]}

%%%%%%%%%%%%%%%%%% Text %%%%%%%%%%%%%%%%%%%%%%

%%%%%%%%%%%%% References %%%%%%%%%%%%%%%%%%%%

% refs with #1=authors, #2=title, #3=publ.ref, #4=hep no :
%\lref\NAME{\paper
%{Authors}{Title(in \it)}{\PLB{No.}{Year}{page},}
%{\hh 0006036 (in\tt)}.}

%\def\hh#1{hep-th/{\it #1}}

% journal~{\bf no.} (year) page

%%%%%%%%%%%%%%%%%%%%%%%%%%%%%%%%%%%%%%%%%%%%%%%%%%%%%%%%%%%%%%%%

\def\a{\alpha} 
\def\b{\beta} 
\def\g{\gamma} 
\def\l{\lambda} 
\def\d{\delta} 
\def\e{\epsilon} 
\def\t{\theta}

\def\L{\Lambda} 
 
\def\p{\partial} 
\def\half{{1\over 2}}

%%%%%%%%%%%%%%%%%%%%%%%%%%%%%%%%%%%%%%%%%%%%%%%%%% 
 
\Title{\vbox{\hbox{YITP-SB-02-14}  
}}  
{\vbox{  
\centerline{On the BRST Cohomology of Superstrings} 
\vskip .2cm 
\centerline{with/without Pure Spinors}}}  
\medskip\centerline{P.A. Grassi$^{~a,}$\foot{pgrassi@insti.physics.sunysb.edu},  
G. Policastro$^{~b,}$\foot{g.policastro@sns.it},  
and  
P. van Nieuwenhuizen$^{~a,}$\foot{vannieu@insti.physics.sunysb.edu}} 
\medskip  
\centerline{$^{(a)}$ {\it C.N. Yang Institute for Theoretical Physics,} } 
\centerline{\it State University of New York at Stony Brook,  
NY 11794-3840, USA} 
\vskip .3cm 
\centerline{$^{(b)}$ {\it Scuola Normale Superiore,} } 
\centerline{\it Piazza dei Cavalieri 7, Pisa, 56126, Italy} 
\medskip 
\vskip  .5cm 
\noindent 
We replace our earlier condition that physical states of the superstring  
have non-negative grading by the requirement that they are analytic in a new real  
commuting constant $t$ which we associate with the central charge of the underlying  
Kac-Moody superalgebra. The analogy with the twisted N=2 SYM theory suggests that  
our covariant superstring is a twisted version of another formulation with an equivariant  
cohomology. We prove that our vertex operators  
correspond in one-to-one fashion to the vertex operators in Berkovits' approach  
based on pure spinors. Also the zero-momentum cohomology is equal in both cases.  
Finally, we apply the methods of equivariant cohomology to the  
superstring, and obtain the same BRST charge as obtained earlier by relaxing the pure spinor  
constraints.  
 
\Date{Jun 19, 2002} 
 
\lref\lrp{ 
U.~Lindstr\"om, M.~Ro\v cek, and P.~van Nieuwenhuizen, in preparation.  
} 
 
\lref\pr{ 
P. van Nieuwenhuizen, in {\it Supergravity `81},  
Proceedings First School on Supergravity, Cambridge University Press, 1982, page 165. 
} 
 
\lref\polc{ 
J.~Polchinski, 
{\it String Theory. Vol. 1: An Introduction To The Bosonic String,} 
{\it String Theory. Vol. 2: Superstring Theory And Beyond,} 
{\it  Cambridge, UK: Univ. Pr. (1998) 531 p}. 
} 
 
\lref\superstring{ 
M.~B.~Green and  J.~H.~Schwarz, {\it Covariant Description Of Superstrings,}  
Phys.\  Lett.\ {\bf B136} (1984) 367; M.~B.~Green and J.~H.~Schwarz, 
  {\it Properties Of The Covariant Formulation Of Superstring Theories,} 
  Nucl.\ Phys.\ {\bf B243} (1984) 285\semi 
M.~B.~Green and C.~M.~Hull, QMC/PH/89-7 
{\it Presented at Texas A and M Mtg. on String Theory, College 
  Station, TX, Mar 13-18, 1989}\semi 
R.~Kallosh and M.~Rakhmanov, Phys.\ Lett.\  {\bf B209} (1988) 233\semi 
%M.~B.~Green and C.~M.~Hull, ``The Brst Cohomology Of An N=1 Superparticle,'' 
%{\it  In *College Station 1990, Proceedings, Strings 90* 133-147. };  
U. ~Lindstr\"om, M.~Ro\v cek, W.~Siegel,  
P.~van Nieuwenhuizen and A.~E.~van de Ven, Phys. Lett. {\bf B224} (1989)  
285, Phys. Lett. {\bf B227}(1989) 87, and Phys. Lett. {\bf B228}(1989) 53;  
S.~J.~Gates, M.~T.~Grisaru, U.~Lindstr\"om, M.~Ro\v cek, W.~Siegel,  
P.~van Nieuwenhuizen and A.~E.~van de Ven, 
{\it Lorentz Covariant Quantization Of The Heterotic Superstring,} 
Phys.\ Lett.\  {\bf B225} (1989) 44;  
A.~Mikovic, M.~Ro\v cek, W.~Siegel, P.~van Nieuwenhuizen, J.~Yamron and 
A.~E.~van de Ven, Phys.\ Lett.\  {\bf B235} (1990) 106;  
U.~Lindstr\"om, M.~Ro\v cek, W.~Siegel, P.~van Nieuwenhuizen and 
A.~E.~van de Ven,  
{\it Construction Of The Covariantly Quantized Heterotic Superstring,} 
Nucl.\ Phys.\  {\bf B330} (1990) 19 \semi 
F. Bastianelli, G. W. Delius and E. Laenen, Phys. \ Lett. \ {\bf 
  B229}, 223 (1989)\semi 
R.~Kallosh, Nucl.\ Phys.\ Proc.\ Suppl.\  {\bf 18B} 
  (1990) 180 \semi 
M.~B.~Green and C.~M.~Hull, Mod.\ Phys.\ Lett.\  {\bf A5} (1990) 1399\semi  
M.~B.~Green and C.~M.~Hull, Nucl.\ Phys.\  {\bf B344} (1990) 115\semi 
F.~Essler, E.~Laenen, W.~Siegel and J.~P.~Yamron, Phys.\ Lett.\  {\bf B254} (1991) 411\semi  
  F.~Essler, M.~Hatsuda, E.~Laenen, W.~Siegel, J.~P.~Yamron, T.~Kimura 
  and A.~Mikovic,  
  Nucl.\ Phys.\  {\bf B364} (1991) 67\semi  
J.~L.~Vazquez-Bello, 
  Int.\ J.\ Mod.\ Phys.\  {\bf A7} (1992) 4583\semi 
E. Bergshoeff, R. Kallosh and A. Van Proeyen, ``Superparticle 
  actions and gauge fixings'', Class.\ Quant.\ Grav {\bf 9}  
  (1992) 321\semi 
C.~M.~Hull and J.~Vazquez-Bello, Nucl.\ Phys.\  {\bf B416}, (1994) 173 [hep-th/9308022]\semi 
P.~A.~Grassi, G.~Policastro and M.~Porrati, 
{\it Covariant quantization of the Brink-Schwarz superparticle,} 
Nucl.\ Phys.\ B {\bf 606}, 380 (2001) 
[arXiv:hep-th/0009239]. 
} 
 
\lref\bv{ 
N. Berkovits and C. Vafa, 
{\it $N=4$ Topological Strings}, Nucl. Phys. B433 (1995) 123,  
hep-th/9407190.} 
 
\lref\fourreview{N. Berkovits,  {\it Covariant Quantization Of 
The Green-Schwarz Superstring In A Calabi-Yau Background,} 
Nucl. Phys. {\bf B431} (1994) 258, ``A New Description Of The Superstring,'' 
Jorge Swieca Summer School 1995, p. 490, hep-th/9604123.} 
 
\lref\oo{ 
%\lref\OoguriPS{ 
H.~Ooguri, J.~Rahmfeld, H.~Robins and J.~Tannenhauser, 
{\it Holography in superspace,} 
JHEP {\bf 0007}, 045 (2000) 
[arXiv:hep-th/0007104]. 
%%CITATION = HEP-TH 0007104;%% 
} 
 
\lref\bvw{ 
N.~Berkovits, C.~Vafa and E.~Witten, 
{\it Conformal field theory of AdS background with Ramond-Ramond flux,} 
JHEP {\bf 9903}, 018 (1999) 
[arXiv:hep-th/9902098]. 
%%CITATION = HEP-TH 9902098;%% 
} 
\lref\wittwi{ 
E.~Witten, 
{\it An Interpretation Of Classical Yang-Mills Theory,} 
Phys.\ Lett.\ B {\bf 77}, 394 (1978);  
E.~Witten, 
{\it Twistor - Like Transform In Ten-Dimensions,} 
Nucl.\ Phys.\ B {\bf 266}, 245 (1986);  
J.~P.~Harnad and S.~Shnider, 
{\it Constraints And Field Equations For Ten-Dimensional  
Super-Yang-Mills Theory,} 
Commun.\ Math.\ Phys.\  {\bf 106}, 183 (1986). 
} 
 
\lref\SYM{ 
W.~Siegel, 
{\it Superfields In Higher Dimensional Space-Time,} 
Phys.\ Lett.\ B {\bf 80}, 220 (1979)\semi 
B.~E.~Nilsson, 
{\it Pure Spinors As Auxiliary Fields In The Ten-Dimensional  
Supersymmetric Yang-Mills Theory,} 
Class.\ Quant.\ Grav.\  {\bf 3}, L41 (1986);  
B.~E.~Nilsson, 
{\it Off-Shell Fields For The Ten-Dimensional Supersymmetric  
Yang-Mills Theory,} GOTEBORG-81-6\semi 
S.~J.~Gates and S.~Vashakidze, 
{\it On D = 10, N=1 Supersymmetry, Superspace Geometry And Superstring Effects,} 
Nucl.\ Phys.\ B {\bf 291}, 172 (1987)\semi 
M.~Cederwall, B.~E.~Nilsson and D.~Tsimpis, 
{\it The structure of maximally supersymmetric Yang-Mills theory:   
Constraining higher-order corrections,} 
JHEP {\bf 0106}, 034 (2001) 
[arXiv:hep-th/0102009];  
M.~Cederwall, B.~E.~Nilsson and D.~Tsimpis, 
{\it D = 10 superYang-Mills at O(alpha**2),} 
JHEP {\bf 0107}, 042 (2001) 
[arXiv:hep-th/0104236]. 
} 
\lref\har{ 
J.~P.~Harnad and S.~Shnider, 
{\it Constraints And Field Equations For Ten-Dimensional  
Super-Yang-Mills Theory,} 
Commun.\ Math.\ Phys.\  {\bf 106}, 183 (1986). 
} 
\lref\wie{ 
%\lref\WiegmannHN{ 
P.~B.~Wiegmann, 
{\it Multivalued Functionals And Geometrical Approach  
For Quantization Of Relativistic Particles And Strings,}  
Nucl.\ Phys.\ B {\bf 323}, 311 (1989). 
%%CITATION = NUPHA,B323,311;%% 
} 
\lref\purespinors{\'E. Cartan, {\it Lecons sur la th\'eorie des spineurs},  
Hermann, Paris (1937)\semi 
C. Chevalley, {\it The algebraic theory of Spinors},  
Columbia Univ. Press., New York\semi 
 R. Penrose and W. Rindler,  
{\it Spinors and Space-Time}, Cambridge Univ. Press, Cambridge (1984)  
\semi 
P. Budinich and A. Trautman, {\it The spinorial chessboard}, Springer,  
New York (1989). 
} 
\lref\coset{ 
P.~Furlan and R.~Raczka, 
{\it Nonlinear Spinor Representations,} 
J.\ Math.\ Phys.\  {\bf 26}, 3021 (1985)\semi 
%%CITATION = JMAPA,26,3021;%% 
A.~S.~Galperin, P.~S.~Howe and K.~S.~Stelle, 
{\it The Superparticle and the Lorentz group,} 
Nucl.\ Phys.\ B {\bf 368}, 248 (1992) 
[arXiv:hep-th/9201020]. 
%%CITATION = HEP-TH 9201020;%% 
} 
 
% 
%\lref\superspace{ 
%S.~J.~Gates, M.~T.~Grisaru, M.~Ro\v cek and W.~Siegel, 
%{\it Superspace, Or One Thousand  
%And One Lessons In Supersymmetry,''} 
%Front.\ Phys.\  {\bf 58}, 1 (1983) 
%[arXiv:hep-th/0108200].} 
% 
\lref\GS{M.B. Green, J.H. Schwarz, and E. Witten, {\it Superstring Theory,}  
 vol. 1, chapter 5 (Cambridge U. Press, 1987).   
} 
\lref\carlip{S. Carlip,  
{\it Heterotic String Path Integrals with the Green-Schwarz  
Covariant Action}, Nucl. Phys. B284 (1987) 365 \semi R. Kallosh,  
{\it Quantization of Green-Schwarz Superstring}, Phys. Lett. B195 (1987) 369.}  
 \lref\john{G. Gilbert and  
D. Johnston, {\it Equivalence of the Kallosh and Carlip Quantizations  
of the Green-Schwarz Action for the Heterotic String}, Phys. Lett. B205  
(1988) 273.}  
\lref\csm{W. Siegel, {\it Classical Superstring Mechanics}, Nucl. Phys. B {\bf 263} (1986)  
93\semi  
W.~Siegel, {\it Randomizing the Superstring}, Phys. Rev. D {\bf 50} (1994), 2799. 
}    
\lref\sok{E. Sokatchev, {\it  
Harmonic Superparticle}, Class. Quant. Grav. 4 (1987) 237\semi  
E.R. Nissimov and S.J. Pacheva, {\it Manifestly Super-Poincar\'e  
Covariant Quantization of the Green-Schwarz Superstring},  
Phys. Lett. B202 (1988) 325\semi  
R. Kallosh and M. Rakhmanov, {\it Covariant Quantization of the  
Green-Schwarz Superstring}, Phys. Lett. B209 (1988) 233.}   
\lref\many{S.J. Gates Jr, M.T. Grisaru,  
U. Lindstrom, M. Ro\v cek, W. Siegel, P. van Nieuwenhuizen and  
A.E. van de Ven, {\it Lorentz-Covariant Quantization of the Heterotic  
Superstring}, Phys. Lett. B225 (1989) 44\semi  
R.E. Kallosh, {\it Covariant Quantization of Type IIA,B  
Green-Schwarz Superstring}, Phys. Lett. B225 (1989) 49\semi  
M.B. Green and C.M. Hull, {\it Covariant Quantum Mechanics of the  
Superstring}, Phys. Lett. B225 (1989) 57.}   
 \lref\fms{D. Friedan, E. Martinec and S. Shenker,  
{\it Conformal Invariance, Supersymmetry and String Theory},  
Nucl. Phys. B271 (1986) 93.} 
\lref\kawai{ 
T.~Kawai, 
{\it Remarks On A Class Of BRST Operators,} 
Phys.\ Lett.\ B {\bf 168}, 355 (1986).} 
 \lref\ufive{N. Berkovits, {\it  
Quantization of the Superstring with Manifest U(5) Super-Poincar\'e  
Invariance}, Phys. Lett. B457 (1999) 94, hep-th/9902099.}   
\lref\BerkovitsRB{ N.~Berkovits,  
{\it Covariant quantization of the superparticle  
using pure spinors,} [hep-th/0105050].   
%%CITATION =HEP-TH 0105050;%%  
}  
%%% berkovits %%%% 
%%% berkovits %%%% 
 
\lref\berko{ 
%\BerkovitsFE 
%\lref\BerkovitsFE{ 
N.~Berkovits, 
{\it Super-Poincar\'e covariant quantization of the superstring,} 
JHEP { 0004}, 018 (2000) 
[hep-th/0001035]%} 
\semi 
%\BerkovitsPH 
%\lref\BerkovitsPH{ 
N.~Berkovits and B.~C.~Vallilo, 
{\it Consistency of super-Poincar\'e covariant superstring tree amplitudes,} 
JHEP { 0007}, 015 (2000) 
[hep-th/0004171]%} 
\semi 
%\BerkovitsNN 
%\lref\BerkovitsNN{ 
N.~Berkovits, 
{\it Cohomology in the pure spinor formalism for the superstring,} 
JHEP { 0009}, 046 (2000) 
[hep-th/0006003]%} 
\semi 
%\BerkovitsWM 
%\lref\BerkovitsWM{ 
N.~Berkovits, 
{\it Covariant quantization of the superstring,} 
Int.\ J.\ Mod.\ Phys.\ A { 16}, 801 (2001) 
[hep-th/0008145]%} 
\semi 
%\BerkovitsYR 
%\lref\BerkovitsYR{ 
N.~Berkovits and O.~Chandia, 
{\it Superstring vertex operators in an AdS(5) x S(5) background,} 
Nucl.\ Phys.\ B {\bf 596}, 185 (2001) 
[hep-th/0009168]; %} 
%\semi 
%\BerkovitsZY 
%\lref\BerkovitsZY{ 
N.~Berkovits, 
{\it The Ten-dimensional Green-Schwarz  
superstring is a twisted Neveu-Schwarz-Ramond string,} 
Nucl.\ Phys.\ B {\bf 420}, 332 (1994) 
[hep-th/9308129] 
%%CITATION = HEP-TH 9308129;%%%} 
\semi 
%\BerkovitsUS 
%\lref\BerkovitsUS{ 
N.~Berkovits, 
{\it Relating the RNS and pure spinor formalisms for the superstring,} 
[hep-th/0104247]; %} 
%\BerkovitsMX 
%\lref\BerkovitsMX{ 
N.~Berkovits and O.~Chandia, 
{\it Lorentz invariance of the pure spinor BRST cohomology  
for the  superstring,} 
[hep-th/0105149]. 
} 
 
%\GrassiUG 
\lref\GrassiUG{ 
P.~A.~Grassi, G.~Policastro, M.~Porrati and P.~Van Nieuwenhuizen, 
{\it Covariant quantization of superstrings without pure spinor constraints},  
[hep-th/0112162]. 
%%CITATION = HEP-TH 0112162;%% 
} 
 
\lref\kacmoody{ 
V.~G.~Kac, {\it Simple graded Lie algebras of finite growth}, Func. Anal. Appl. {\bf 1},  
(1967) 328;  
K.~Bardacki and M.~B.~Halpern, {\it New dual quark model}, Phys. Rev. D {\bf 3} (1971) 2493;  
V.~G.~Kac, {\it Infinite dimensional Lie algebras}, 3rd edition, Cambridge University Press,  
Cambridge, 1990. 
} 
 
%\BerkovitsUE 
\lref\BerkovitsUE{ 
N.~Berkovits and P.~Howe, 
{\it  
Ten-dimensional supergravity constraints from the pure spinor formalism   
for the superstring}, [hep-th/0112160]. 
%%CITATION = HEP-TH 0112160;%% 
} 
 
%\WittenZZ 
\lref\WittenZZ{ 
E.~Witten, 
{\it Mirror manifolds and topological field theory,} 
hep-th/9112056. 
%%CITATION = HEP-TH 9112056;%% 
} 
 
\lref\wichen{ 
E.~Witten, 
{\it Chern-Simons gauge theory as a string theory,} 
arXiv:hep-th/9207094. 
%%CITATION = HEP-TH 9207094;%% 
} 
 
%\lref\bgi{ 
%C.~Becchi, S.~Giusto and C.~Imbimbo, 
%{\it The holomorphic anomaly of topological strings,} 
%Fortsch.\ Phys.\  {\bf 47}, 195 (1999) 
%[hep-th/9801100]\semi 
%C.~Becchi, S.~Giusto and C.~Imbimbo,  
%{\it Topological B models}, unpublished.  
%} 
 
%\howe 
\lref\howe{P.S. Howe, {\it Pure Spinor Lines in Superspace and  
Ten-Dimensional Supersymmetric Theories},  
Phys. Lett. B {\bf 258} (1991) 141, Addendum-ibid.B259 (1991) 511\semi  
P.S. Howe, {\it Pure Spinors, Function Superspaces and Supergravity  
Theories in Ten Dimensions and Eleven Dimensions}, Phys. Lett. B {\bf 273} (1991)  
90.}

%\KazamaQP 
\lref\Kazama{ 
Y.~Kazama and H.~Suzuki, 
{\it New N=2 Superconformal Field Theories And Superstring Compactification,} 
Nucl.\ Phys.\ B {\bf 321}, 232 (1989). 
%%CITATION = NUPHA,B321,232;%% 
%\Figueroa-O'FarrillPV 
%\lref\Figueroa-O'FarrillPV{ 
J.~M.~Figueroa-O'Farrill and S.~Stanciu, 
%``N=1 and N=2 cosets from gauged supersymmetric WZW models,'' 
arXiv:hep-th/9511229. 
%%CITATION = HEP-TH 9511229;%% 
} 
 
\lref\BilalRN{ 
A.~Bilal and J.~L.~Gervais, 
{\it BRST Analysis Of Super Kac-Moody And Superconformal Current Algebras,} 
Phys.\ Lett.\ B {\bf 177}, 313 (1986).} 
 
\lref\GrassiTZ{ 
P.~A.~Grassi, G.~Policastro and P.~van~Nieuwenhuizen, 
{\it The massless spectrum of covariant superstrings,} 
arXiv:hep-th/0202123.} 
 
\lref\equivariant{ 
J.~M.~F.~Labastida and M.~Pernici, Phys. Lett. {\bf B 212}, 56 (1988)\semi 
S.~Ouvry, R.~Stora and P.~van Baal, 
{\it On The Algebraic Characterization Of Witten's Topological Yang-Mills Theory,} 
Phys.\ Lett.\ B {\bf 220}, 159 (1989)\semi 
L.~Baulieu and I.~M.~Singer, 
{\it Conformally Invariant Gauge Fixed Actions For 2-D Topological Gravity,} 
Commun.\ Math.\ Phys.\  {\bf 135}, 253 (1991)\semi 
J.~Kalkman, 
{\it BRST Model For Equivariant Cohomology And Representatives For The Equivariant Thom Class,}  
Commun.\ Math.\ Phys.\  {\bf 153}, 447 (1993)\semi 
R.~Stora, 
{\it Exercises in equivariant cohomology,} 
CERN-TH-96-279 
{\it NATO Advanced Study Institute: Quantum Fields and  
Quantum Space Time, Cargese, Corsica, France, 22 Jul - 3 Aug 1996}\semi 
R.~Stora, 
{\it Exercises in equivariant cohomology,} arXiv:hep-th/9611114 
}

\lref\concoho{ 
A.~Blasi and R.~Collina, 
{\it Basic Cohomology Of Topological Quantum Field Theories,} 
Phys.\ Lett.\ B {\bf 222}, 419 (1989)\semi 
F.~Delduc, N.~Maggiore, O.~Piguet and S.~Wolf, 
{\it Note on constrained cohomology,} 
Phys.\ Lett.\ B {\bf 385}, 132 (1996) 
[arXiv:hep-th/9605158]} 
 
\lref\sorella{ 
F.~Fucito, A.~Tanzini, L.~C.~Vilar, O.~S.~Ventura, C.~A.~Sasaki and S.~P.~Sorella, 
{\it Algebraic renormalization: Perturbative twisted considerations on  
topological Yang-Mills theory and on N = 2 supersymmetric gauge 
theories,} arXiv:hep-th/9707209.} 
 
\lref\swcp{ 
S.~Cordes, G.~W.~Moore and S.~Ramgoolam, 
{\it Lectures on 2-d Yang-Mills theory, equivariant cohomology and topological  
field theories,} 
Nucl.\ Phys.\ Proc.\ Suppl.\  {\bf 41}, 184 (1995) 
[arXiv:hep-th/9411210]\semi 
M.~Dubois-Violette, 
{\it  
A Bigraded version of the Weil algebra and of the Weil homomorphism for  
Donaldson invariants,} 
J.\ Geom.\ Phys.\  {\bf 19}, 18 (1996) 
[arXiv:hep-th/9402063]\semi 
L.~Baulieu, A.~Losev and N.~Nekrasov, 
{\it Chern-Simons and twisted supersymmetry in various dimensions,} 
Nucl.\ Phys.\ B {\bf 522}, 82 (1998) 
[arXiv:hep-th/9707174]. 
} 
 
\lref\donwitt{S. Donaldson, J. Diff. Geom. {\bf 18} 269 (1983);  
J. Diff. Geom. {\bf 18} 397 (1987) \semi 
E. Witten, Comm. Math. Phys. {\bf 117} 353 (1988)} 
 
\lref\henne{M.~Henneaux, {\it BRST cohomology  of  the fermionic string},   
Phys.~Lett.~B {\bf 183} 59, (1987).}  
 
\lref\ciccio{W.~Siegel,   
{\it Boundary conditions in first quantization}  
Int.~J.~Mod.~Phys.~A {\bf 6}, 3997, (1991)} 
 
\lref\bigpicture{N.~Berkovits, M.~T.~Hatsuda and W.~Siegel, 
{\it The Big picture}, Nucl.\ Phys.\ B {\bf 371}, 434 (1992) [arXiv:hep-th/9108021].}

%%%%%%%%%%%%%%%%%%%%%%%%%%%%%%%%%%%%%%%%%%%%%%%%%% 
\baselineskip14pt 
 
\newsec{Introduction} 
 
Recently, we developed a new approach to  
the long-standing problem of the covariant quantization of the superstring \GrassiUG.  
The formulation of  Berkovits of the super-Poincar\'e covariant superstring  
in $9+1$ dimensions \berko~ is based on a {\it free} conformal field theory  
on the world-sheet and a 
nilpotent BRST charge which defines the physical vertices as 
representatives of its cohomology. In addition to the conventional 
variables $x^m$ and $\t^\a$ of the Green-Schwarz formalism, a 
conjugate momentum $p_\a$ for $\t^\a$ and a set of commuting ghost fields 
$\l^\a$ are introduced. The latter are complex Weyl spinors satisfying the 
pure spinor conditions $\l^\a \g^m_{\a\b} \l^\b = 0$ (cf. for example \howe).  
This equation can be solved by decomposing $\lambda$ with 
respect to a non-compact $U(5)$ subgroup of $SO(9,1)$ into a singlet 
$\underline{1}$, a vector $\underline{5}$, and a tensor 
$\underline{10}$. The vector can be expressed in terms of the singlet 
and tensor, hence there are 11 independent complex variables in 
$\l^\a$. 
 
Since the presence of the non-linear constraint $\l^\a \g^m_{\a\b}\l^\b = 0$  
makes the theory unsuitable for a path integral quantization and higher loop computations,  
we relaxed the pure spinor condition by adding further ghosts. We were naturally led to a finite 
set of extra fields, but the BRST charge $Q$ of this system was not 
nilpotent, and the central charge of the conformal field theory did 
not vanish. The latter problem was solved by adding one more extra 
ghost system, which we denoted by $\eta^m$ and $\omega^m_z$. The 
former problem was solved by introducing yet another new ghost pair, 
$b$ and $c_z$, which we tentatively associated with the central charge 
generator in the affine superalgebra which plays an essential role in the 
superstring \csm. 
 
The BRST charge is linear in $c_z$, and without further conditions on physical states the theory  
would be trivial. We proposed that physical states belong not only to the BRST cohomology 
($ Q \, |\psi \rangle = 0$, but $ |\psi \rangle \neq Q\, |\phi\rangle$), but also that  
the deformed stress tensor $T+ {\cal V}^{(0)}$, where ${\cal V}^{(0)}$ is a vertex operator,  
satisfies the usual OPE of a conformal spin 2 tensor. (The latter condition is weaker that the  
requirement that vertex operators be primary fields with conformal spin equal to 1).   
 
The definition proposed in \GrassiTZ~replaced the stress tensor condition by the  
requirement that the physical states belong to a subspace ${\cal H}'$  
of  the entire linear space ${\cal H}$ of vertex operator. The latter can be decomposed  
w.r.t. a grading naturally associated with the underlying affine algebra as ${\cal H} = {\cal H}_- \oplus {\cal H}_+ $,  
with negative and non-negative grading, respectively. The  BRST charge $Q = \sum_{n\geq 0} Q_n $ contains only  
terms $Q_n$ with non-negative grading, hence one can consistently consider the action of  
$Q$ in ${\cal H}_+$. The physical space is identified with the cohomology group  $H(Q, {\cal H}_+)$,  
namely  
\eqn\leb{\eqalign{ &  Q |\psi \rangle = 0\,, ~~~~~~~~~~  
|\psi \rangle~ \in {\cal H}_+ \,, \cr 
& |\psi \rangle \neq Q\, |\phi \rangle\,, ~~~~~~ |\phi \rangle~ \in {\cal H}_+\,. 
}} 
 
Furthermore, by rescaling the ghost fields with a parameter $t$ to the power equal to the grading  
of the ghost field and assigning the grading $-1/2$ to the parameter $t$,  
we restate the definition of physical states as the BRST cohomology of vertices with  
vanishing grading and analytical in the new parameter $t$.  
 
The essential point is that the cohomology in the pure-spinor formulation \berko~is a constrained  
cohomology and this translates in our formalism into an equivariant cohomology.  
This implies that the physical observables are identified  
not naively by the BRST cohomology, but with the classes of an equivariant cohomology.  
This is evident from the structure of our BRST operator and from the fact that  
on the complete functional space the BRST cohomology is trivial. Usually,  
in that situation one has to identify what is the functional space on which the BRST  
cohomology should be computed and, depending on the context, one has to determine  
an operator which defines such physical states.  
 
At the time when we completed paper \GrassiTZ~we were not aware of the fact that  
the functional subspace characterized by the non-negative graded monomials  
was indeed the subspace on which the BRST cohomology becomes an equivariant  
cohomology, but we did observe that it gives the correct spectrum for the superstrings.  
In the present paper, we  completely spell out the equivalence  
between the grading condition and the equivariant cohomology. 
  
We also want to mention that the same situation can be found in the  
context of topological Yang-Mills, topological sigma models and RNS  
superstrings \henne. Essentially, also in those cases the BRST cohomology is not  
well defined due to the commuting character of superghosts  
unless a further condition is imposed. For example in \ciccio, to avoid the  
ambiguities of the cohomology in presence of commuting ghosts, Siegel introduced non-minimal  
terms in the action and observed that suitable combinations of fields and constraints can  
be read as creation and annihilation operators acting in Hilbert space. The definition of the  
vacuum removes the ambiguities in the cohomology computations.\foot{Similar problems occur if one 
adds a BRST invariant field $Y_\a$ such that $Y_\a \l^\a \neq 0$ (see for example 
\lref\tonin{
I.~Oda and M.~Tonin,
%``On the Berkovits covariant quantization of GS superstring,''
Phys.\ Lett.\ B {\bf 520}, 398 (2001)
[arXiv:hep-th/0109051]; 
M.~Matone, L.~Mazzucato, I.~Oda, D.~Sorokin and M.~Tonin,
%``The superembedding origin of the Berkovits pure spinor covariant  quantization of superstrings,''
arXiv:hep-th/0206104.}
\tonin), but they can be solved by using our grading (restricted to $\l^\a$) and 
our  definition of physical states.} 
 
The paper is organized as follows: in section 2, we review the definition of the grading and   
of the decomposition of the BRST charge according to it. In section 3, we  
restate the condition on physical states and we show how the BRST charge presented in \GrassiUG~can be  
reformulated in the context of the equivariant cohomology. This leads to the same result achieved in  
\GrassiUG, but the interpretation is different. In section 4, as a pedagogical example and to  
underline the relation between the present formulation with the equivariant cohomology theories, we  
review the Donaldson-Witten model in $D=(4,0)$ and the relation with the twisted $N=2$ SYM. In section 5,  
we present a proof of the equivalence of the pure-spinor cohomology with our formulation and some  
examples. In section 6, we reproduce the results of \GrassiUG~starting from yet another point of view, but  
which illustrates some of the details in the proof of the previous section. In section 7, as a last  
application, we compute the zero momentum cohomology.  
 
\newsec{Grading} 
 
Following \GrassiUG, we review the definition of the grading, the construction of its  
worldsheet current and the decomposition of the BRST charge according to the grading.  
 
We have based our approach on the following affine superalgebra \csm  
\eqn\dope{\eqalign{ 
&d_\a(z) d_\b(w) \sim -{{\g^m_{\a\b}\Pi_m(w)}\over{z-w}},\quad \quad\quad\quad 
d_\a(z) \Pi^m(w) \sim{{\g^m_{\a\b}\p\t^\b(w)} \over {z-w}}, \cr 
&\Pi^m(z) \Pi^n(w) \sim- {1\over (z-w)^2} \, \eta^{mn} \, k \,, \quad ~ 
d_\a(z) \p_w \t^\b(w) \sim{1\over (z-w)^2} \, \delta^{~\b}_{\a}\, k\,, \cr  
& \Pi^m(z) \p_w \t^\b(w) \sim 0\,, ~~~~~~~~~~~~~~~~\quad\quad  
\p_z \t^\a(z) \p_w \t^\b(w) \sim 0\,, 
}}  
where $\sim$ denotes the singular contributions to the OPE's.  
 
This algebra has a natural grading defined as follows:  
$d_\a(z)$ has grading $1/2$, $\Pi^m(z)$ has grading $1$,  
$\p_z \theta^\a(z)$ has grading $3/2$, and the central charge $k$ (which  
numerically is equal to unity) has grading $2$. The corresponding ghost systems are  
$(\l^\a, \b_{z \a})$, $(\xi^m, \b_{z m})$, $(\chi_\a, \kappa^{\a}_z)$, and $(c_z, b)$. We thus  
define the following grading for the ghosts and corresponding antighosts  
\eqn\gr{\eqalign{ 
& {\rm gr}(\l^\a) = {1\over 2}\,,~~ \quad {\rm gr}(\xi^m) = {1}\,, \quad 
~~~{\rm gr}(\chi_\a) = {3\over 2}\,,~~ \quad {\rm gr}(c_z) = {2}\,, \cr 
& {\rm gr}(\b_\a) = -{1\over 2}\,, \quad {\rm gr}(\b_m) = {-1}\,, \quad 
{\rm gr}(\kappa^\a) = -{3\over 2}\,, \quad {\rm gr}(b) =-{2}\,. 
}}  
We also need the ghost $\omega^m$ and the antighost $\eta^m_z$,  
although this pair does not seem to correspond to a generator.  
We assign the grading  
${\rm gr}(\eta^m_z) = - 2$ and ${\rm gr}(\omega^m) = 2$ for the following  
reason. In \GrassiUG, we relaxed the pure spinor constraint by  
successively adding quartets starting from $(\l_+, \l_{[ab]}; \b^+, \b^{[ab]})$ of  
\berko~(the indices $a,b$ belong to the fundamental representation of the $U(5)$  
subgroup of $SO(1,9)$), and adding the fields  
$(\l^a, \b_a; \xi^a, \b'_a)$ with grading $(1/2,-1/2,1,-1)$. This procedure yields the covariant  
spinors $\l^\a$ and $\b_{\a}$, but now the fields $(\xi^a, \b'_a)$ are  
non-covariant w.r.t. $SO(9,1)$. Thus, we added the quartet  
$(\xi_a, \b^{'a}; \chi_a, \kappa^a)$ with grading  $(1,-1, 3/2,-3/2)$. The spinors  
$(\chi_a, \kappa^a)$ are part of a covariant spinor and the missing parts are  
introduced by adding the quartets $(\chi^+, \kappa_+; c, b)$   
and $(\chi^{[ab]}, \kappa_{[ab]}; \omega^m, \eta^m)$, both with  
grading $(3/2,-3/2, 2,-2)$. In this way, we obtain the covariant fields $\l^\a = (\l_+, \l^a, \l_{[ab]})$;  
$\b_\a = (\b^+, \b_a, \b^{[ab]})$; $\xi^m = (\xi^a, \xi_a)$; $\b^m = (\b^{'a}, \b'_a)$;  
$\chi_\a =(\chi^+, \chi_a, \chi^{[ab]})$; $\kappa^\a= (\kappa_+, \kappa^a, \kappa_{ab})$;  
$b,c$ and $\eta_{m} ,\omega^m$.  
 
As usual for a conformal field theory, it is natural to introduce a current  
whose OPE's with the ghost and antighosts reproduce the grading assignments in \gr 
\eqn\currgrad{ 
j_z^{grad} = - {1\over 2} \b_{z,\a} \l^\a - \b_{z,m} \xi^m - {3\over 2} \k^\a_{z} \chi_\a - 2 \, b\, c_z  
- 2 \, \eta^m_z \omega_{m}\,. 
} 
Independent confirmation that this current might be important  
comes from the cancellation of the anomaly (namely the terms with $(z-w)^{-3}$) in  
the OPE of the stress energy tensor $T_{zz}(z)$  
(cf. eqs.~(1-3) of ref.~\GrassiUG) with $j_z^{grad}$. In fact, one finds  
\eqn\graano{ 
c^{grad} =   {1\over 2}\times ( + 16)_{\l\b} + 1 \times (-10)_{\xi\b} +   
{3\over 2} \times (+16)_{\kappa\chi} + 
2\times (-1)_{bc} + 2 \times (-10)_{\eta\omega} = 0\,. 
} 
The requirement that the vertex operators contain only terms with non-negative grading  
leads to the correct massless spectrum \GrassiTZ. It will also  
severely restrict the contribution of the vertex operators to correlation functions  
(in the usual RNS approach ghost insertions are needed to compensate the anomaly in the ghost  
current, whereas here we anticipate to need insertions of fields in   
${\cal H}_-$ to compensate the  
non-negative grading of vertex operators ${\cal U}^{(1)} \, \in {\cal H}_+$).  
 
All the terms in the stress tensor $T_{zz}(z)$ and in the ghost current  
\eqn\ghostfinal{\eqalign 
{ 
T_{zz} &= -\half \Pi^m_z  \Pi_{m z} - d_{z \a} \p_z \t^\a - \b_{z m} \p_z \xi^m -  
\b_{z \a} \p_z \l^\a - \k^\a_z \p_z \chi_\a + \p_z b\, c_z -   
\eta^m_z  \p_z \, \omega_m \,, \cr 
J^{gh}_z &= - \left(  
\beta_{m z} \xi^m + \kappa_z^\a \chi_\a + \beta_{z \a} \l^\a + b\, c_z  
+ \eta^m_z \omega_{m} \right)   \,, 
}} 
have grading zero, since they are sums of terms of ghost and antighost pairs with  
opposite grading. On the other hand, the terms in the  
 current $j^B_z(z)$ (cf. eq.~(1.2) in \GrassiUG) and the field $B_{zz}(z)$  
have different grading\foot{In \GrassiUG~we presented four different solutions $B^{i}$ of the  
the equation $T_{zz}(z) = \{ Q, B^{i}_{zz}(z)\}$. None of the solutions $B^{i}$ have definite  
grading except $B^{IV}_{zz}(z)  =  
b \, \hat{T}_{zz}(z)  + b\p_z b c_z - \half \p^2_z b$ which has grading equal to  
$-2$ carried by the antighost $b$.}. For instance, the BRST current can be  
decomposed into the following pieces $j^B_z(z) = \sum_{n=0}^2 j^{B,(n)}_z(z)$ 
\eqn\decBRST{\eqalign{ 
j^{B,(0)}_z(z) &= -\, \xi^m \kappa^\a_z \g_{m \a\b} \l^\b  
-  {1\over 2} \l^\a \g^m_{\a\b} \l^\b \b_{z m} + \cr 
& - \half b \left( \xi^m \p_z \xi_m - {3\over 2} \chi_\a \p_z \l^\a + \half \p_z \chi_\a \l^\a \right)  
- {1\over 2} \p_z \left( b\, \chi_\a \l^\a \right) \,, \cr 
  j^{B,({1 \over 2})}_z(z) &= \l^\a d_{z \a} \,,  
~~~~~~~~~~~~~~~j^{B,({1})}_z(z) = - \xi^m \Pi_{z m} \,, \cr 
 j^{B,({3 \over 2})}_z(z) &= - \chi_\a \p_z \theta^\a \,, ~~~~~~~~~~~j^{B,({2})}_z(z) = c_z \,.       
}}  
It is clear that all terms in $j^B_z(z)$ have non-negative grading.  
 
\newsec{A New Definition of Physical States.} 
 
We begin with some notation that will be used in the following. We  
denote the quantities in Berkovits' formalism  with pure spinor constraints  
with a lower index $B$. For example $Q_B$ is his BRST charge and ${\cal U}^{(1)}_B$ is his  
unintegrated vertex operator. The physical spectrum of superstrings 
is identified with the ghost number $+1$ elements of the cohomology  
$H(Q_B|{\cal H}_{p.s.})$ where ${\cal H}_{p.s.}$ is the linear vector space of vertex  
operators expressed as polynomials of the world-sheet fields $x^m, \theta^\a$ and of  
the pure spinors $\lambda^\a$. The latter satisfy the pure spinor condition $\l \g_m \l=0$.  
The group $H(Q_B|{\cal H}_{p.s.})$ is an example of a constrained BRST cohomology, or  
equivalently, of equivariant cohomology \equivariant. In the latter case, the BRST cohomology  
is computed on the supermanifold $x^m, \theta^\a$ on which the space-time  
translations $ x^m \rightarrow x^m + \half \lambda \g^m \lambda $, generated by  
unconstrained spinors $\l^\a$, act freely. One finds that $Q^2_B = - {\cal L}_{V}$ where  
$V^m= \half \lambda \g^m \lambda$. (In Howe's work on pure  
spinors \howe~ a translation $x^m \rightarrow x^m + \l\g^m \bar\l$  is considered where $\l$ are  
pure spinors. The integrability condition for a covariantly constant field, $\l^\a \nabla_\a \phi =0$  
lead to the SYM field equations). 
 
In order to compare with our formalism \GrassiUG, let us rescale the pure spinors with  
a constant commuting parameter $t \in I\!\!R$. One  
can interpret this constant as the quartic root of the central charge of the  
Kac-Moody algebra, $t^4 = k$. Using the gradings discussed in the  
previous section, we obtain $Q_B = \oint \l^a d_\a \rightarrow  
t \oint \l^a d_\a$ and $Q^2_B = t^2 \oint \half \lambda \g^m \lambda \, \Pi^m$.  
Notice that the r.h.s. can be also written in term of the Lie derivative  
${\cal L}_V = d\, {\iota}_V +   {\iota}_V \, d$, where $  {\iota}_V$ is the  
contraction of a form with the vector $V^m$. One can represent $\iota_V$ by the  
operator $\oint dz \, V^m \b_{zm}$; its action on (parity reversed forms) $\xi^m$ is then  
given by the OPE of $\b_{zm}(z)$ with $\xi^m(z)$. 
The exterior differential $d$ is $\xi^m \p_m$ where $\xi^m$  
are the parity-reversed coordinates of the cotagential bundle $\Pi T^*{\cal M}$.  
The usual exterior derivative $d = dx^m \p_m $ has been replaced by  
$- \oint dz\, \xi^m \Pi_{zm}$ . Since $\Pi^m_z(z) \p^l x^n(w) \sim (z-w)^{-l-1}$, the  
operator $- \oint dz\, \xi^m \Pi_{zm}$ represents the exterior derivative on the jet bundle  
$\{x^m, \p\, x^m, \p^2 x^m, \dots\}$. One may represent $\Pi_z^m$ by the functional derivative  
${\d / \d x^m(z)}$, but note that the latter operator has a central charge  
and the former has not. The definition in terms of $\Pi^m_z$  
is explicitly supersymmetric.  
Following the approach of equivariant cohomology \equivariant,  
one can define a new BRST operator $Q'$ by  
\eqn\newBRST{ 
Q' =t\, Q_B + t^2 d + \iota_V =  
 Q - t^ 2  \oint \xi^m \Pi_{z m} - \oint {1\over 2} \l^\a \g^m_{\a\b} \l^\b \b_{z m}\,. 
} 
Unfortunately, this operator fails to be nilpotent for two reasons: the operator $d$  
does not commute with $Q_B$ and $d^2 \neq 0$.  
Notice that this is a quantum effect: in fact the Kac-Moody generator $d_\a$  
on the space of functions on the superspace ${\cal M}$ acts like the  
covariant derivative $\oint d_\a \, F(x,\t) = D_\a F(x,\t)$, and in the  
same way $\Pi_m$ acts like the ordinary space-time derivative. This is  
clearly true only on functions on the superspace ${\cal M}$ and  
not on forms of $\Omega^*({\cal M})$. In addition, one has to take into  
account that the OPE of $\Pi^m$ with itself has a central term.  
Computing the square of $ Q'$ one finds 
\eqn\nil{\eqalign{ 
(Q')^2 &= t^2\, \Big( Q_B^2 +  \,  d\, \iota_V + \iota_V\, d  \Big)  +  
t^3 \, \{ Q_B, d \}  + t^4 d^2 \cr 
&= t^3 \oint \xi_m  \l^\a \g^m_{\a\b} \p_z \theta^\b + t^4 \oint \xi^m \p_z \xi_m \,. 
}} 
where we used $Q^2_B = - {\cal L}_{V}$ from \berko~and we also  
used that $Q_B = \oint \l^\a d_\a$ anticommutes with $\iota_V$, and  
$\iota_V$ anticommutes with itself. According to the grading of \GrassiTZ,  
$\xi_m \g^m_{\a\b}\l^\b$ has grading $3/2$, and we associate the factor $t^3$ to  
$\p_z \theta^\b$ because then the whole expression for $(Q')^2$ gets grading $-3/2$  
(we define ${\rm gr}(t) =-1/2$).  
  
The $t^3$ term generates fermionic translations in the extended  
superspace ${\cal M}'$ parametrized by the coordinates $(x^m,\t^\a,\phi_\a)$ and   
constructed in \csm. However, as noticed by Siegel,  
since $\{ i {\p \over \p \phi_\a}, i {\p \over \p \phi_\b} \} = 0$,  
one can apply the first order constraint $ i {\p \over \p \phi_\a} =0$ to eliminate  
the variable $\phi_\a$, obtaining the usual superspace ${\cal M}$. Since  
$ \p_z \theta^\b$ generates translations of the variable $\phi_\a$,  
we can view it again as a Lie derivative and repeat the construction in \newBRST. 
Namely, the first term in \nil~can be seen as a Lie derivative ${\cal L}_\psi$  
along the fiber $\phi_\a$ of the superspace  
${\cal M}'$ with respect the spinor $\psi_\a =\xi_m \g^m_{\a\b} \l^\b$. We  
have  
\eqn\newnewBRST{ 
Q'' = Q' + t^3 d_{\phi} + \iota_\psi  
} 
where $d_{\phi} = \oint \chi_\a  \p_z \theta^\a$  and  
$\iota_\psi = - \oint \k^\a \xi_m \g^m_{\a\b} \l^\b$.  
One can again square this expression and study the terms on the right hand side. One finds  
only terms proportional to $t^4$, and these terms are $Q''$ invariant. At first sight they seem  
not to contain any new translation generator. However, adding $c_z(z)$ plus $b(z)$  
time the $t^4$ terms yields the final BRST charge (\GrassiUG,\kawai).  
It coincides with the expression we derived in \GrassiUG 
\eqn\BRSTfinall{\eqalign{ 
Q_0 &=  - \oint \Big( \xi^m \kappa^\a_z \g_{m \a\b} \l^\b  
+ {1\over 2} \l^\a \g^m_{\a\b} \l^\b \b_{z m} + 
 \half b \, ( \xi^m \p_z \xi_m - {3\over 2} \chi_\a \p_z 
\l^\a + \half \p_z \chi_\a \l^\a ) \Big) \,, \cr 
Q & =  
t \oint  \l^\a d_{z \a} - t^ 2 \oint \xi^m \Pi_{z m} -  
t^3 \oint \chi_\a \p_z \theta^\a + t^4 \oint c_z  + Q_0 \,. 
}} 
First, we note that the BRST  charge $Q$ is a polynomial in the  
constant $t$ and the ghost terms collected in $Q_0$  
are $t$-independent. As a consequence $Q^2_0=0$\foot{In \BilalRN~a non-nilpotent $Q$  
has been found, but it contained a $Q_0$ which is nilpotent.  
This $Q_0$ corresponds to our $Q_0$.}.  
This is a well-known fact in the Sugawara construction based on a  
super-Kac-Moody \Kazama.  
Extracting the ghosts $(\l^a,\xi^m,\chi_\a)$ from $Q$ and $Q_0$ leads to two  
representation of the generators of the same affine  
algebra, namely $(d_\a, \Pi^m, \p \t^\a)$ and  
$\Big( (-\b_m \g^m \l - \xi_m \g^m \k - b \, \p \chi - 3/4\, \p b \, \chi)_\a,  
-\k \g^m \l - b \, \p \xi^m - 1/2 \,\p b \, \xi^m, b \, \p \l^\a + 1/4 \, \p b \, \l^\a)\Big)$.  
Next, we note that by assigning the grading  
to the fields discussed before and  
the grading $-1/2$ to the parameter $t$, the BRST charge obtains zero grading\foot{ 
In the case of topological field theories obtained from supersymmetric models 
by twisting, the grading corresponds to the ${\cal R}$-charge \sorella.}. 
Since the parameter $t$ is constant the assignment of this grading does not spoil  
the cancellation of the anomaly of the grading current.  
It is interesting to compute the BRST transformations of the antighosts 
\eqn\ACTd{\eqalign{ 
&\{ Q, b\} = t^4 \,, \cr 
&[Q, \k^\a_z] = - t^3 \p_z \t^\a   + b\, \p_z \l^\a + {1\over 4} \left(  \p_z b\right)  \l^\a \,, \cr 
&\{ Q, \b^m_z \} = - t^2 \Pi^m_z - \k_z \g^m \l + b \, \p_z \xi^m + \half 
\left( \p_z b \right) \xi^m \,, \cr  
&[Q, \b_{z \a}] = t\, d_{z\a} - \b^m_{z} (\g_m \l)_\a - \xi^m (\g_m \k_z)_\a - b \, \p_z \chi_\a  
- {3\over 4} \left( \p_z b \right) \chi_\a\,. 
}}  
From the $t$-dependent terms it becomes  
evident that the BRST transformation of $b$ contains the central charge of the  
Kac-Moody algebra. Being a number, one can set it to $1$. We refer to \sorella~(remark 17  
on page 48) for a discussion of this point 
 
The BRST charge $Q$, the stress tensor $T_{zz}$,   
the ghost current $J^{gh}_z$ and the action $S$ (see \GrassiUG) have grading zero. Thus,  
we require that physical observables have zero grading as well.  
A generic vertex operator ${\cal U}$ can be expanded into power series of the parameter $t$,  
$ {\cal U} = \sum_{n = - N_-}^{N_+} t^n \, {\cal U}_{n}$ where $N_-$ and $N_+$ are  
the lowest and the highest power of $t$. In general $N_- \geq 0$, and  
the numbers $N_-$ and $N_+$ are bounded for a fixed ghost number  
and at fixed world-sheet conformal weight  
(the latter is number of $z$ indices in the expression for the unintegrated vertex ${\cal U}$).  
The definition of physical states presented in \GrassiTZ~can be now reformulated by  
requiring that  {\it the vertex operators are analytic functions of $t$}, as earlier  
proposed for topological gauge theories \sorella~and for topological sigma models  
\concoho. This is  
completely equivalent to our previous requirement that only ${\cal H}_+$ 
with non-negative graded operators has to be taken into account \GrassiTZ. In the following, the  
space ${\cal H}_+$ is identified with the analytic functions of the parameter  
$t$. To justify the choice of functional space, we note that  
\eqn\cohoCC{ 
Q_B = \lim_{t \rightarrow 0} t^{-1} Q(t)\,,   
} 
if $\xi_m = \chi_\a = c = \l \g^m \l =0$, namely if all additional ghost fields (except the pure spinors)  
are set to zero and $\l$ satisfy the pure spinor constraints.  
 
As an example, we  consider the vertex operator massless  
states in the open string  
\eqn\cohoC{\eqalign{ 
{\cal U}^{(1)}(z) &= t\, \l^\a A_\a + t^2\, \xi^m A_m + t^3\,  
\chi_\a W^\a  + t^4\, \omega^m B_m \cr 
& + b\, \Big( {1\over t^{2}} \l^\a \l^\b F_{\a\b} + {1\over t} 
\l^\a \xi^m F_{\a m} + t^0 \xi^m \xi^n F_{m n}   \cr 
& + t^0 \l^\a \chi_\b F^{~~\b}_\a + t \, \chi_\a \, \xi^m F^{\a}_{~~m} + t^2  
\chi_\a \chi_\b F^{\a \b} \Big) \cr  
& + b\, \omega^m \Big( t \, \lambda^\a G_{m\a} + t^2 \xi^n G_{mn} +  
t^3 \chi_\a G_{m}^{~~\a} \Big) + t^4 
b\, \omega^m \omega^n K_{mn} \,,  
}} 
where $ A_\a, \dots, K_{mn}$ are arbitrary superfields of $x_m, \t^\a$. The  
analyticity w.r.t.  $t$ implies that the first two terms in the first  
bracket should be canceled. The rest of the vertex is polynomial in $t$ and  
\eqn\inj{ 
\lim_{t \rightarrow 0}\left. t^{-1} {\cal U}^{(1)}(z)\right|_{\xi_m = \chi_\a = c =0}  
=  {\cal U}^{(1)}_B(z) \,, 
}  
namely it coincides with pure-spinor unintegrated vertex.  
%Notice that $b=0$ is consistent with the  
%condition $t=0$ since $\{Q, b\} = t^4$.  
In fact, by identifying $t = k^{1\over 4}$, where $k$ is the Kac-Moody central  
charge, setting $k=t=0$, implies that the OPE of $\Pi^m$ with itself vanishes, and  
the BRST charge is consequently nilpotent. There is a caveat in this argument:  
$\left. \p_t Q\right|_{t = b =0}^2 \neq 0$ as we know from  
\berko. But if $\l^\a$ satisfies the pure spinor constraint, it is nilpotent.  This  
point will be clarified in the forthcoming sections.  
 
\newsec{N=2  D=4 SYM and Topological Yang-Mills} 
 
The introduction of grading by means of a constant parameter $t$ and  
the requirement that the space of unintegrated vertex operator be restricted to  
non-negative grading or to analytical functions of $t$ is a common  
situation in so-called equivariant cohomology theories \equivariant.  We believe that our  
covariant superstring is related to a worldsheet supersymmetric model by a suitable  
twisting. It may be illuminating to review the relation between  
N=2 SYM in D=(4,0) dimensions and the topological Donaldson-Witten model \donwitt~because  
these models are also related by twisting, and the cohomology after the twisting is also restricted  
to the polynomials which are analytical in the constant twisted supersymmetry ghost $t$.  
 
The N=2 supersymmetric theory is described by a gauge potential 
$A_\m$, the gauginos $\psi^i_\a, \bar\psi^i_{\dot\a}$ and a complex 
scalar $\phi$. The index $i =1,2$ is the index of the R-symmetry group 
$U(2)$. The subgroup $U(1)$ determines the $R$-charge. All fields 
carry an index $a$ in the adjoint representation of the gauge group which 
we suppress. By twisting the R-symmetry with one of the $SU(2)$ 
subgroup of $SO(4)$, one obtains fermions with Lorentz-vector 
indices, and the susy parameters become a Lorentz-scalar $t$, a vector and a 
self-dual antisymmetric tensor 
\eqn\twistA{ 
\psi_\mu = \bar\sigma^{i \dot\a}_\mu \psi_{i \dot\a}\,, ~~~~~  
\chi_{\m\n} = \sigma_{\m\n}^{i \a} \psi_{i \a}\,, ~~~~~ 
\eta = \e^{i \a} \psi_{i \a}\,, } 
$$ 
\e_\mu = \bar\sigma^{i \dot\a}_\mu \zeta_{i \dot\a}\,, ~~~~~  
t_{\m\n} = \sigma_{\m\n}^{i \a} \zeta_{i \a}\,, ~~~~~ 
t =  \e^{i \a} \zeta_{i \a}\,. 
$$ 
With the gauge potential $A_\mu$ and the complex scalar $\phi$, these are the fields of the  
Donaldson-Witten model. To compare the fields of the two different models, the Wess-Zumino  
gauge has been chosen in superspace, and susy auxiliary fields have been eliminated.  
In this particular case, the susy transformations generated by $ q^i_\a$ and $\bar q^i_{\dot\a}$  
close only up to gauge transformations and up to equation of motions  
\eqn\twistB{ 
\{ q^i_\a, \bar q_{j \dot\a} \} =  
\delta^i_{~j} \sigma^\m_{\a\dot\a} \p_\mu  + {\rm gauge ~transf.} + {\rm eqs. ~of ~motion}\,, 
} 
$$ 
\{ q^i_\a, q^j_{\b} \} =     \{ \bar q^i_{\dot\a}, \bar q^j_{\dot\b} \} =     
{\rm gauge ~transf.} + {\rm eqs. ~of ~motion} \,.  
$$ 
 
To define the supersymmetric and gauge invariant observables in the N=2 susy model, one  
needs to define a new BRST operator which is the sum of the usual BRST operator $Q$,  
the supersymmetry generators and the translation generator multiplied by their constant ghosts  
(the commuting $\zeta^{i\a}$ and $\bar\zeta^{i\dot\a}$ and the anticommuting   
$\tau^\mu$) and a further term 
\eqn\twistC{ 
Q_{S} = Q + \zeta^{i \a} q_{i\a} + \bar\zeta^{i \dot\a} \bar q_{i\dot\a} +  
\tau^\mu \p_\mu 
- \zeta^{i \a} \sigma^\mu_{\a\dot\a}  \bar\zeta_i^{\dot\a} \partial_{\tau^\m} \,. 
} 
The last term is needed in order to make $Q_S$ nilpotent on all classical fields and on ghost except  
the gauginos. Further, $Q$ contains also terms which transform the Yang-Mills ghost $c^a$  
into two supersymmetry ghosts $\zeta^{i\a}$ and $\bar\zeta^{i\dot\a}$.  
Nilpotency of $Q_S$ on the gauginos can be achieved by adding to the theory suitable antifields and  
constructing the corresponding BRST operator of the BV formalism.  
 
Twisting the supersymmetry generators,  
we find Witten's fermionic symmetry $\d_W = \e^{i \a} \, q_{i \a}$, the  
vector supersymmetry $\delta_\mu = \bar\sigma^{i \dot\a}_\mu \, \bar q_{i \dot\a}$ and  
the self-dual antisymmetric tensor supersymmetry $\delta_{\mu\nu} = \bar\sigma^{i \a}_{\mu\nu} \, q_{i \a}$.   
The corresponding BRST operator is given by  
\eqn\twistC{ 
Q_{T} = Q + t\,  \d_W + \e^\mu \d_\mu + \tau^\mu \p_\mu  
- t\, \e^\m \partial_{\tau^\m} \,, 
} 
where the ghost $t$ is associated to $\d_W$ and $\e^\mu$ to $\delta_\m$. $Q_{T}$ is again nilpotent  
on all fields except the selfdual antisymmetric tensor $\chi_{\mu\nu}$. We  
drop the antisymmetric generator $\delta_{\mu\nu}$  
since the observables are completely determined by the remaining symmetries. By twisting the  
fields of the supersymmetric action, the new fields will carry the same $R$-charge as  
before twisting and in  
particular $t$ carries the charge $-1$. Explicitly, the transformations generated by  $Q_{T}$  
are given by (we can set $\e^\m=\tau^\mu =0$ without affecting the conclusions\foot{By this we mean  
that $\e^\mu$ and $\tau^\mu$ do not transform into terms without either $\e^\mu$ or $\tau^\mu$, implying  
that we can apply filtration methods.}) 
\eqn\twistD{ 
[Q_{T} , A_\mu] = - \nabla_\mu c + t\, \psi_\mu\,, ~~~~~~~~~~~~ 
\{Q_{T} , \psi^\m\} = \{ c, \psi^m \} - t\, \nabla_\mu \phi\,,  
} 
$$ 
\{Q_T, c\} = c^2 - t^2\, \phi\,, ~~~~~ 
[Q_T , \phi] = [c, \phi]\,, ~~~~~ 
[Q_T , \bar \phi] =[ c, \bar\phi] + 2 \, t\, \eta\,,  
$$ 
$$ 
\{Q_T, \eta\} = \{ c, \eta\} + {t \over 2} [\phi, \bar\phi]\,, ~~~~ 
\{Q_T ,\chi_{\mu\nu}\} = \{ c, \chi_{\mu\nu} \} + t\, F^+_{\mu\nu} + {t^2 \over 2} \chi^*_{\mu\nu}\,, 
$$ 
$$ 
[Q_T ,\chi^*_{\mu\nu}] = -2\, (\nabla_{[\mu} \psi_{\nu]})^+ + 2\, [\phi, \chi_{\mu\nu}] + [c, \chi^*_{\mu\nu}]\,,    
$$ 
where $\chi^*_{\mu\nu}$ is the antifield of $\chi_{\mu\nu}$. Here the superscript $+$ denotes  
the selfdual part of the tensor. For the purposes of the  
present section we will not describe the action of $Q_T$ on the antifields. It can be shown that   
the cohomology of $Q_T$ is independent from the antifields \lref\maggiore{F. Delduc, N. Maggiore,  
O. Piguet and S. Wolf, {\it Note on constrained cohomology}, Phys. Lett. {\bf B385}, 132 (1996)}  
\maggiore.  
 
The crucial point is that the cohomology of $Q_T$ is only non-trivial if one restricts the  
space of polynomials to those which are analytical  in the  
global ghosts $t,\e^\m$ and $t^\m$ \maggiore. In fact, the cohomological classes are  
generated by monomials ${\cal P}_n(\phi)$  of the undifferentiated fields $\phi$ 
\eqn\twistE{ 
{\cal P}_n(\phi) = {1\over n} tr\Big( \phi^n\Big)\,, ~~~~~ n \geq 2\,. 
} 
Thus the cohomology is not only restricted to monomials analytic in $t$, but it is even independent of $t$.  
Due to the commuting nature of $\phi$, the expressions $ tr\Big( \phi^n\Big)$ for $n$ sufficiently  
large is related to higher order Casimir invariants of the gauge group.  
 
The analysis of the proof in \maggiore~is based on a filtration of the functional  
space (which contains the constant ghosts $t, \e^\mu,\tau^\m$), and of the BRST operator  
with respect to the counting operator $N = t \p_t$. One has  $Q_T = \sum_{n=0}^2 Q_n$, where  
\eqn\twistF{ 
Q^2_0 = 0\,, ~~~~ \{Q_1, Q_0\} = 0\,, ~~~~ 
Q_1^2 + \{ Q_0, Q_1\} =0\,,~~ 
\{Q_1, Q_2\} = 0\,, ~~~ 
Q^2_2 =0\,.} 
The first term of the decomposition $Q_0$ selects the pure gauge transformations in the  
BRST symmetry \twistD~whereas $Q_1$ and  $Q_2$ lead to shift transformations.\foot{In the  
case of superstrings, the charge $Q_0$ in \BRSTfinall~implements the pure spinor constraint at the  
level of cohomology (it generates the gauge transformations of the antighost fields).  
The charge $Q-Q_0$ in \BRSTfinall~leads to shifts of the fields as in the topological model.}  
 
By relaxing the constraint of analyticity, it is easy to show that all monomials  ${\cal P}_n(\phi) $ become 
BRST trivial. For instance we have  
\eqn\twistG{ 
tr\Big( \phi^2 \Big) = \left\{ Q, tr \Big( - {1 \over t^{2}} c\, \phi + {  1\over 3 t^4 } c^3 \Big) \right\} \,.  
} 
In other words, working in the functional space whose elements are power series  
in the global ghosts (in particular $t$), namely ${\cal U} = \sum_{n\geq0} t^n {\cal U}_n$, the  
cohomology is non-trivial, but in the larger space with also negative powers of $t$ the BRST  
cohomology becomes trivial, in agreement with the results of Labastida-Pernici and  
Baulieu-Singer \equivariant.\foot{It is interesting to note that in the  
string case, by imposing the restriction that ${\cal U} = \sum_{n\geq1} t^n {\cal U}_n$.  
The cohomology is further restricted to the states of topological super-Yang-Mills in  
D=(9,1). This might lead to the construction of topological super-Yang-Mills model in higher  
dimensions where the action is given by  
$S = \langle \Psi, Q_T \Psi \rangle$. Clearly, one needs a definition of the inner product in order  
to have a gauge invariant and supersymmetric model.} 
 
In terms of the cohomological representatives \twistE~, one  
can construct the solution to the descent equations:  
$\{ Q, \Omega^{n}_{p} \} + d\, \Omega^{n+1}_{p-1} = 0$, where  
$d$ is the exterior differential and $\Omega^{n}_{p}$ are $p$-forms  
with ghost number $n$. The generators of the equivariant  
cohomology of $Q_T$ satisfy the descent equations  
\eqn\descent{  
\eqalign{ 
&[Q_T,  {1\over 2 t^4} tr F^2] =\, - d\, {1\over t^3} tr\Big( F\,\psi \Big)\,, ~~~~~~ 
\{Q_T , {1\over t^3} tr \Big(F\,\psi\Big) \} =\,  - d\,  {1\over t^2} tr \bigl(\phi\, F + \half \psi^2\bigr)\,,\cr 
& 
[Q_T , {1\over t^2} tr\bigl(\phi\, F + \half \psi^2\bigr)] =\, - d\,  tr \Big( {1 \over t} \phi \psi \Big) \,, ~~~~~ 
\{Q_T , tr \Big( {1 \over t} \phi \psi \Big) \} = -\half d\,\Tr \phi^2\,, \cr  
&[Q_T , \half tr \phi^2] =\, 0\,.} 
}  
Except the last element of the descent equations, namely the monomial $tr \phi^2$,  
all the other generators are explicitly non-analytical. The same situation will happen  
in the  case of open superstrings: the descent equations are given by  
$\{Q, {\cal U}\} = 0$ and $[Q, {\cal V}_z] = \p_z {\cal U}$.  
Here ${\cal U}$ corresponds to the so-called non-integrated vertex and  
${\cal V}_z$ to the integrand of the integrated vertex. Following the suggestions of  
topological models, one finds that ${\cal U}$ is written in terms of a power series of $t$, but  
${\cal V}_z$ will contain also non-analytical terms. It turns out that those non-analytical  
pieces are irrelevant for computations of amplitudes.  
 
\newsec{Equivalence with Berkovits' formulation} 
 
In the case of massless states a direct comparison with  
the equation of motions obtained in \berko~can be easily  
done, but, for massive states, the field equations  
in N=1 d=(9,1) superspace formulation are not known. Only recently, the  
equations of motion for the first massive state for open superstring has been  
derived in 
\lref\BerkovitsQX{N.~Berkovits and O.~Chandia, 
{\it Massive superstring vertex operator in D = 10 superspace,} 
arXiv:hep-th/0204121.}  
\BerkovitsQX~using the pure spinor formulation.  
 
Since the comohology $H^{(1)}(Q_B|{\cal H}_{p.s.})$  
has been proved in \berko~to contain uniquely the spectrum of the  
RNS superstring, or equivalently of the Green-Schwarz string quantized in  
the light-cone gauge, it will be sufficient to prove the equivalence of our cohomology group  
$H^{(1)}(Q, {\cal H}_+)$ with the pure spinor constrained cohomology   $H^{(1)}(Q_B|{\cal H}_{p.s.})$.  
 
Both the BRST operator \BRSTfinall~and the vertex operators are 
analytic functions of an indeterminate variable $t$. We are therefore 
studying a cohomology with values in a ring of analytic functions of 
$t$. However, as discussed in \lref\Atiyah{M.~F.~Atiyah and R.~Bott, 
{\it The Moment Map and Equivariant Cohomology}, Topology, {\bf 23}, 1(1984)}  
\Atiyah, we can work at a fixed value $t=t_0$ as long as the 
multiplication by the monomial $(t- t_0)$ is an injective map in the 
cohomology.  In our case, the presence of a grading implies that this 
is true for any value of $t$ except possibly for $t=0$.  In fact, the 
equation $(t-t_0) \,{\cal U} =0$ can be separated according to the 
grading in $t \, {\cal U} =0$ and $t_0 \, {\cal U} =0$.  The latter is 
only satisfied for ${\cal U} =0$ unless $t_0=0$.  This means that, in 
analyzing the cohomology, we can consider $t$ as a given non-zero real 
parameter. 
 
The next step is to prove that the cohomology is in fact independent of the value of $t$. This follows from the  
fact that one can change the value of $t$ by applying a similarity transformation to the BRST operator.  
More precisely, defining $Q_{grad}$ to be the grading charge, $Q_{grad} = \oint j_z^{grad}$, one has the following  
``evolution'' equation  
\eqn\evolution{  
t {\partial \over \partial t } Q(t) = [Q_{grad}, Q(t)] .}   
This equation is in fact the statement that $Q(t)$ is an homogeneous function of grading zero in $t$ and all the  
fields. Since $Q_{grad}$ does not depend on $t$, the equation is easily solved by  
$Q(t) = e^{Q_{grad} \ln {t \over t_1}} Q(t_1) e^{- Q_{grad} \ln {t \over t_1}}$.  
Thus $Q(t_1)$ is related to $Q(t)$ by a similarity transformation that is, however, singular at $t=0$.  
Following the ideas of Witten~  
\lref\WittenIM{ 
E.~Witten, {\it Supersymmetry And Morse Theory}, 
J.\ Diff.\ Geom.\  {\bf 17}, 661 (1982). 
}\WittenIM,  
we will consider the limit $t \to 0$. For this purpose, it 
is more convenient to use the operator $D_t \equiv {1 \over t} Q(t)$, 
which has of course the same cohomology as $Q$. In eq. \evolution, the 
left-hand side is manifestly at least linear in $t$, but the 
right-hand side is also linear because the $t$-independent term $Q_0$ 
commutes with $Q_{grad}$. We can then divide both sides by $t$ and get 
$\partial_t D_t = [Q_{grad}, D_t]$. The main idea of our proof is that, 
in the limit $t\to 0$, the divergent term in $D_t$, $Q_0 /t$, has the 
effect of localizing the cohomology on the fixed points of the action 
of $Q_0$. The transformation properties of various fields under $Q_0$ 
are given below: 
\eqn\fixedpoints{\eqalign{ 
&\{Q_0, \xi^m \} = - \half \l \g^m \l \,,\cr 
&\{Q_0, \chi_\a \} = \xi^m (\g_m \l)_\a \,,\cr 
& \{Q_0, c \} = \xi^m \p \xi_m + \l^\a \p\chi_\a - \chi_\a \p \l^\a \,,} 
} 
one can see that the fixed points of $Q_0$ are $\l \g^m \l =0, \,\xi^m =0, \,\chi_\a =0$ and $c_z=0$.  
The first of this conditions, of course, reproduces the pure-spinor constraints, and the other ghosts are set to zero. The  
BRST operator reduces to the $Q_1$ term, that reproduces the Berkovits' one, and $Q_0$. The only difference with  
Berkovits' cohomology is that the vertices can still depend on $\beta_{zm}$, the antighost of $\xi^m$.  
We must recall that the pure-spinor constraint $\l \g^m \l =0$ implies that the antighost of $\l$  
has the gauge-invariance $w_\a \to w_\a + \L_m (\g^m \l)_\a$,   
for an arbitrary parameter $\L_m$. The vertex operators must then be restricted to be invariant with respect to this  
gauge transformations \BerkovitsQX. In our formalism we do not see this requirement even after the localization.   
But we must still consider the action of $Q_0$, that exactly reproduces the transformations:  
$\{Q_0, w_\a\} = \beta_m (\g^m \l)_\a$.  Vertices that are not gauge-invariant are ruled out by the cohomology of $Q_0$.  
This completes the proof of the equivalence of our cohomology with the pure-spinor one.   
 
As an illustration of this point, we consider explicitly the first massive level of the open superstring. After the localization,  
the most general form of the vertex, at ghost number 1, is  
\eqn\vertex{\eqalign{ 
{\cal U}_z^{(1)} =& \p \l^\a A_\a(x,\t) + \p \t^\b \l^\a B_{\a\b}(x,\t) + d_\b \l^\a C^\b_\a (x,\t) + \Pi^m \l^\a H_{m\a}(x,\t) + \cr 
& + w_\a \l^\b \l^\g E^\a_{(\b\g)}(x,\t) + \b_m \l^\b \l^\g F^m_{(\b\g)}(x,\t) \,.  
}} 
Comparing with the vertex in \BerkovitsQX, one can see that the only difference is in the  
second line, where the second term is absent and the first one only appears in the gauge-invariant combinations  
$J  \l^\a E_\a$ and $N^{mn} \l^\a E_{[mn]\a}$, where $J = w_\a \l^\a$ and $N^{mn} = w_\a (\g^{mn})^\a_\b \l^\b$.  
Requiring that $Q_0$ annihilates the vertex implies $\b_m \l^\b \l^\g \l^\d \g^m_{\a(\b} E^\a_{\g\d)} = 0$ and  
$\k^\a \l^\b \l^\g \l^\d \g_{m \a(\b} F^m_{\g\d)} =0 $.  
The coefficient $E^\a_{(\g\d)}$, considered as a matrix  
in the indices $\a,\g$, can be expanded on a basis of Dirac matrices, and the  
expansion contains terms with 0, 2 or 4 gamma matrices. The terms with 0 and 2 matrices reproduce the Berkovits'  
terms. The term with 4 matrices has to satisfy $  \b_m \l^\a \l^\b \l^\g \g^{[mpqrs]}_{(\a\b}  E_{\g) [pqrs]} = 0$ and  
by decomposing $\l^\a \l^\g \rightarrow \l \g^{tuvwx}\l \g^{\a\g}_{tuvwx}$, one obtains the  
equation $ \g^{[mpqrs]}_{\b \a} \g^{\a\g}_{tuvwx} E_{\g [pqrs]} = 0$ which implies that $ E_{\g [pqrs]} = 0$.  
On the other hand, decomposing  $F^m_{(\b\g)}$ as a 5-form  $F^m_{[npqrs]} \g^{[npqrs]}_{\a\b}$, one immediately  
obtains that $F^m_{[npqrs]} = 0$.  
 
At this point, to construct the elements of the cohomology for $t \neq 0$, it is convenient to  
disentangle the vertices and the BRST charge into fixed grading numbers. We shall show that  
only four equations must be really solved: all the others give only algebraic  
relations among the different pieces of the vertices and can be easily solved.  
  
As already mentioned, the BRST charge $Q$ is an analytic functions of $t$ up to power  
four: $Q = \sum_{n=0}^4 t^n Q_n$ (in order to simplify the notation, we denote $Q_B$ by  $Q_1$).  
The nilpotency of $Q$ is translated into the relations 
\eqn\speB{ 
\sum_{n=0}^{m} \{ Q_{m-n}, Q_n \} = 0\,, ~~~~~ m=0,\dots,8\,, ~~~~~~~ Q_{n}=0\,,~~~ n>4\,. 
} 
However, due to the particular form of the various $Q_n$, the equations  
\speB~reduce to 
\eqn\speC{\eqalign{ 
&Q_0^2 =0\,, \quad\quad \! \!\! \{ Q_0, Q_B \} =0\,,  \quad\quad Q_B^2 + \{ Q_0, Q_2 \} =0\,, \cr 
&\{ Q_0, Q_3 \} + \{ Q_2, Q_B \} =0\,, \quad\quad \,  Q_2^2 + \{ Q_B, Q_3 \} +  
\{ Q_0, Q_4 \} =0 \,,\cr  
&  \{ Q_2, Q_3 \} =0\,,   \quad\quad \! Q_3^2 = 0\,,   
\quad\quad \{Q_i, Q_4\} =0\,, ~~~ i =1,\dots,4\,, 
}}  
A generic vertex operator ${\cal U}^{(1)}$ for the open superstring with ghost number $1$ 
belongs to ${\cal H}_+$ and it  
can be expressed in terms of a power series of the parameter $t$,  
${\cal U}^{(1)} = \sum_{n\geq 0} t^n \, {\cal U}_n$.  
This implies that expanding the equation $\{ Q, {\cal U}^{(1)} \}= 0$ in different powers  
we have the following equations 
\eqn\specD{\eqalign{ 
& \{ Q_0, {\cal U}_0\} = 0\,, \cr 
& \{ Q_0, {\cal U}_1\} + \{ Q_B, {\cal U}_0\}  = 0\,, \cr 
& \{ Q_0, {\cal U}_2\} + \{ Q_B, {\cal U}_1\} +  \{ Q_2, {\cal U}_0\} = 0\,,  \cr 
& \{ Q_0, {\cal U}_3\} + \{ Q_B, {\cal U}_2\} +  \{ Q_2, {\cal U}_1\} + \{ Q_3, {\cal U}_0\} = 0\,, \cr 
&  \{ Q_0, {\cal U}_n\} + \{ Q_B, {\cal U}_{n-1}\} +  \{ Q_2, {\cal U}_{n-2}\} +  
\{ Q_3, {\cal U}_{n-3}\} + \{ Q_4, {\cal U}_{n-4}\}= 0\,, ~~~ n \geq 4 \,. 
}} 
Using the fact that $b^2 =0$, we can decompose any contribution ${\cal U}_n$ into  
a $b$-dependent term and a $b$-independent one, ${\cal U}_n = {\cal U}'_n + b \Delta_{n}$.  
We therefore decompose all the equations into a $b$-dependent part and a $b$-independent  
one. Since $\{Q_i, b \} =0$ for $i=1,\dots,3$, and $\{ Q_4 , b \} =1$, the $b$-independent  
equations for  
$n \geq 4$ can be easily solved. For example, let us consider the  
equation for $n=4$; we can solve it for $\Delta_0$  
\eqn\specE{ 
- \Delta_0 =  
\{ Q_0, {\cal U}'_4\} + \{ Q_B, {\cal U}'_3\} +  \{ Q_2, {\cal U}'_2\} + \{ Q_3, {\cal U}'_1\}\,. 
} 
In a similar way all $\Delta_n$ with $n>0$ are solved by using \specD~with $n>4$. Note that  
$\{ Q, {\cal U} \} =0$ can be decomposed into $Q = Q' + Q_4$ and $(Q')^2 + \{ Q', Q_4\} =0$ and  
${\cal U} = {\cal U}' + b\, \Delta$, this implies $\{Q', {\cal U}'\} + \Delta=0$ and $\{Q', \Delta\} =0$  
(as a consequence of \specD). Now, inserting $\Delta = - \{ Q', {\cal U}' \}$ in $\{Q', \Delta\} =  
- \{ \{ Q', Q'\}, {\cal U}' \} = \{\{ Q', Q_4\} , {\cal U}' \} =0$ and $\{Q', \{Q_4, {\cal U} \}\} +  
 \{Q_4, \{Q', {\cal U} \}\} =0$, but $\{Q_4 , {\cal U} \} = \{ Q', \Delta \}$ and $\{Q', {\cal U} \} = - \Delta$  
which is $Q_4$ invariant.   
This fixes all the $\Delta_n$. 
However, from the first equation of \specD~, one gets the  
two equations $\{Q_0, {\cal U}'_0\} = 0$ and $\{ Q_0, \Delta_0\} =0$. The second is  
a constraint on $\Delta_0$ and the solution \specE~should be compatible with it. This can easily  
be proved  by using the  commutation relations \speC~and equations \specD~ 
for ${\cal U}'_i, i=1,\dots,3$. In the same way, one can solve all the equations  
for $n >4$ and the four remaining equations can be now expressed in terms  
of only the $b$-independent part of ${\cal U}_n$. Hence, at this point all the  
equations in $\specD$ for $n\geq 4$ have been solved.  
 
As an example, we illustrate the construction in the case of massless  
vertex for the open superstring. This example will also provide some hints for  
constructing the massive states in the present formalism. 
 
In the massless case, we consider only worldsheet scalar vertex operators. This  
implies that only the antighost $b$ is allowed in the expression for the vertex.  
Moreover, this also implies that ${\cal U}^{(1)} = \sum_{n=0}^{3} t^n {\cal U}^{(1)}_{n}$.  
Now, using the decomposition ${\cal U}^{(1)}_{n} = {\cal U}'^{(1)}_{n} + b \, {\Delta}^{(2)}_{n}$  
and by noting that ${\cal U}'^{(1)}_{0}$ vanishes we can simplify eqs. \specD. 
For $n=4,\dots,7$ we have 
\eqn\massA{\eqalign{ 
&n=4:~~~~~\quad\{Q_B, {\cal U}'^{(1)}_{3} \} + \{Q_2, {\cal U}'^{(1)}_{2} \} + \{Q_3, {\cal U}'^{(1)}_{1} \} + 
\Delta^{(2)}_0 = 0\,, \cr 
&n=5:~~~~~\quad \{Q_2, {\cal U}'^{(1)}_{3} \} + \{Q_3, {\cal U}'^{(1)}_{2} \} + 
\Delta^{(2)}_1 = 0 \,,\cr 
&n=6:~~~~~\quad \{Q_3, {\cal U}'^{(1)}_{3} \} + \Delta^{(2)}_2 = 0\,, \cr 
&n=7:~~~~~\quad \Delta^{(2)}_3 = \,. 0  
}} 
Observing that $\{Q_3, {\cal U}'^{(1)}_{i} \} = 0$ for $i=1,2,3$ because the massless  
vertex ${\cal U}^{(1)}$ cannot depend upon $d_{z\a}$ (and upon the corresponding right-movers  
in the closed string case), we obtain $ \Delta^{(2)}_3 = \Delta^{(2)}_2 = 0$. The remaining  
$\Delta^{(2)}_0$ and $\Delta^{(2)}_1$ depend only upon the variations of ${\cal U}'^{(1)}_{i}$  
with $i=1,2,3$. Moreover, $\Delta^{(2)}_0$ and $\Delta^{(2)}_1$ should satisfy the following  
consistency conditions 
\eqn\massB{\eqalign{ 
&\{Q_0, \Delta^{(2)}_0 \} =0\,, \quad\quad  \quad\quad \quad\quad  
\quad\quad\{ Q_1, \Delta^{(2)}_0 \} + \{ Q_0, \Delta^{(2)}_1 \} =0\,, \cr 
&\{ Q_1, \Delta^{(2)}_1 \} + \{ Q_2, \Delta^{(2)}_0 \} =0\,,  
~\quad\quad \{ Q_2, \Delta^{(2)}_1 \} =0\,, 
}} 
where we have already used $\{Q_3, \Delta^{(2)}_i\} =0$ for $i=0,1$. The $b$-independent  
terms ${\cal U}'^{(1)}_{i}$ with $i=1,2,3$ should satisfy the equations 
\eqn\massC{\eqalign{ 
&\{Q_0, {\cal U}'^{(1)}_{1} \} =0\,,  \cr 
&\{Q_1,  {\cal U}'^{(1)}_{1} \} + \{Q_0,  {\cal U}'^{(1)}_{2} \}  =0\,, \cr 
&\{Q_2,  {\cal U}'^{(1)}_{1} \} + \{Q_1,  {\cal U}'^{(1)}_{2} \}   
+\{Q_0,  {\cal U}'^{(1)}_{23} \}  =0\,. 
}} 
 
From Lorentz invariance, ghost number and analyticity, we have that  
${\cal U}'^{(1)}_{1} = \lambda^\a A_\a(x,\t)$ where $ A_\a(x,\t)$ is a generic superfield.  
It automatically satisfies  the first equation of \massC. Furthermore, we  
have that ${\cal U}'^{(1)}_{2}= \xi^m A_m(x,\t)$ solves the second equation if  
the superfields $ A_\a(x,\t)$ and $A_m(x,\t)$ satisfy  
\eqn\massD{ 
A_m = {1\over 8} \g^{\a\b}_m D_\a A_\b\,, \quad\quad\quad\quad  
\g^{\a\b}_{mnrpq} D_\a A_\b = 0\,.  
}   
The third equation is solved by assuming ${\cal U}'^{(1)}_{3}= \chi_\a W^\a(x,\t)$ if  
the superfield $W^\a$ is related to $ A_\a(x,\t)$ and $A_m(x,\t)$ by the usual  
equation $W^\a = {1\over 10} \g^{\a\b}_m ( D_\b A^m - \p^m A_\a)$. 
  
From eqs. \massA, we have  
\eqn\massE{ 
\Delta_0^{(2)} = - \l^\a \chi_\b D_\a W^\b - \xi^m \xi^n F_{mn} \,,  \quad\quad\quad  
\Delta_1^{(2)} = - \xi^m \chi_\a \p_m W^\a\,, 
} 
where $F_{mn} = {1\over 2} (\p_m A-n - \p_n A_m)$.  
It is easy to verify that the equations \massB~hold because the superfields  
$A_\a, A_m$ and $W^\a$ satisfy  
\eqn\massF{ 
F_{mn} = \g_{mn,\b}^{~~\a} D_\a W^\b\,, ~~~~ D_\a F_{mn} = \left( \g_{[m} \p_{n]} W\right)_\a\,.  
} 
This concludes the example for the massless vertex operator. The result coincides with  
that obtained in \berko~and in \GrassiTZ.  
 
In order to underline again the relevance of the analyticity (or of the grading) to  
select the correct physical spectrum, one can notice that at a given mass level\foot{In the following  
formulae, we denote by the subscript $z_1 \dots z_n$ the conformal weight $n$, in order  
not to be confused with the grading index $p$ of the vertex ${\cal U}_{p}$}  
(or, equivalently, at a given conformal weight) one has the following structure  
for the vertex operators 
\eqn\pcA{ 
{\cal U}_{z_1 \dots z_n} = \sum^{l \leq n}_{i=0} \sum_{(\{p_i\}, \{k_i\})} \prod_{j=0}^i  
\left( \p^{p_i}_z b \right)^{k_1} {\cal U}^{(\{p_i\} \{k_i\})}_{z_1 \dots z_{n-l}} \,\,, 
} 
where $\sum_i p_i = l$, $0 < p_1 < \dots < p_l$ and $k_i = 0,1$.  
For example the first massive vertex operator can be decomposed into  
\eqn\pcB{ 
{\cal U}_z =   {\cal U}^{(\{0,0\}, \{0,0\})}_z +   
b\, {\cal U}^{(\{1,0\}, \{0,0\})}_z  + \p_z b\,  {\cal U}^{(\{0,1\}, \{0,1\})}  +  
b\, \p_z b\,  {\cal U}^{(\{1,1\}, \{0,1\})}\,. 
} 
Since $\{ Q, \p^k_z b \} = 0$, given a vertex operator of a lower level,  
fox example, the massless vertex ${\cal U}$, one can construct an element of the  
BRST cohomology at the next level by $\p_z b\, {\cal U}$. In the same way, at the  
conformal weight $2$ level, one can have $\p^2_z b\,{\cal U}$. This phenomenon is  
unwanted since the total cohomology at a given mass level is not described by a  
single vertex operator. However, the minimum grading of the unwanted terms is $-4, -8, \dots$  
and therefore they are excluded, by choosing analytical (or, positive gradings) vertex  
operators.  
 
Notice that relaxing the constraint on analyticity, one can find the massless vertex as a part  
of the massive vertex operator by selecting the $-4$ grading part of the vertex. In order, to project  
out the unwanted terms one can multiply the vertex operator ${\cal U}_{z_1 \dots z_n}$  
by $\p_z b\, \p^2_z b\, \p^3_z b\, \dots\p^{n}_z b$.

\newsec{Construction of a nilpotent covariant BRST charge based on equivariant cohomology.}  
 
In this section we present yet another derivation of our BRST charge, in addition  
to the derivation in \GrassiUG~based on relaxing constraints by adding new ghosts, and the  
derivation of Sec. 3 based the BRST charge with operators $d$ and $\iota_V$ but not imposing  
the pure spinor constraints.  We assume that the pure spinor constraints is imposed  
each time after performing the OPE's. This is different viewpoint of equivariant cohomology,  
but it clarifies the construction of vertex operators in our cohomology starting  
from those constructed in the pure spinor formulation.  
 
We assume for the present discussion that the spinors $\l^\a$ satisfy  
the pure spinor constraint: $\l \g^m \l =0$. As shown in \berko, in order to  
match correctly the degrees of freedom and to cancel the central charge  
also the conjugate momentum $\b_\a$ describe only 22 dof. This can be achieved by  
observing that the action is invariant under the symmetry $\d\, \b_\a = \L_m (\g^m \l)_\a$ and  
$\d X^m = \d d_\a = \d \t^\a = \d \l^a =0$. The gauge parameters $ \L_m$ removes  
10 dof from $ \b_\a$ matching the corresponding 22 dof of $\l^\a$.  
 
This symmetry is encoded in the BRST $Q_B = \oint \l^a d_\a$,  
by acting twice o $\b_\a$ 
\eqn\appA{ 
\{Q^2_B, \b_\a \}= \{ Q_B, d_\a\} = - \Pi_m (\g^m \l)_\a \,.   
} 
This implies that, on ${\cal H}_{p.s.}$, the BRST charge is nilpotent up to  
gauge transformations with $\L_m = \Pi_m$. However, to study its cohomology,  
it convenient to modify the BRST operator such that it squares to zero on ${\cal H}_{p.s.}$.  
This is equivalent to doing the standard Weyl complex procedure \swcp.  
 
This can be done by extending the BRST transformation of $\b_\a$ by adding  
a gauge transformation  
\eqn\appB{ 
\{ Q'_B, \b_\a\} = \{ Q_B, \b_\a\} - \b_m (\g^m \l)_\a\,,~~~~~ \{ Q_B, {\rm other ~fields} \} = 0\,, 
} 
This approach leads to introduce a  new field $\b_m$ with ghost number $-1$.  On all the  
other fields the action of $Q'_B$ is the same of that of $Q_B$.  
 
The requirement of nilpotency implies that $ ( \{ Q'_B, \b_m\} + \Pi_m ) (\g^m \l)_\a =0$.  
The most general solution of this equation is given by  
$\{ Q'_B, \b_m\} = - \Pi_m - \k^\a \g^m_{\a\b} \l^\b$  
where $\k^\a$ is a new spinorial field. In this way,   
the BRST charge is nilpotent, except on $\b_m$.  
Requiring that $Q'_B$ is nilpotent,  
\eqn\appC{ 
\{Q'_B, \k^\a_z \} = - \p_z \t^\a - b\, \p_z \l^\a - h_z\, \l^a\,,   
} 
where $b$ is a scalar with ghost number $-1$, and $h_z$ is a 1-form also with ghost number  
$-1$. Notice that the new terms are allowed because of the pure spinor condition. Finally,  
imposing that $Q'_B$ is nilpotent on $\k^\a$, we obtain that $\{ Q_B, b \} =1$ and  
$\{Q_B, h \} =0$. A particular realization of $h_z$ is $h = x \, \p b$ where  $x$ is a constant.   
At this point we obtain the same BRST charge as obtained earlier by other methods.   
 
The BRST transformations obtained for the antighost fields $\b_\a, \b_m, \k^\a$ and $b$ coincide  
with \ACTd~if we set  $\xi_m = \chi_\a$ to zero. In particular, we note that this is implied  
by the eq.~\inj~which relates the pure-spinor vertices with those of our cohomology.  
 
The BRST transformation of the field $b$ would render the BRST cohomology  
trivial, if we did not introduce further constraint to define physical states.  
Therefore, we restrict the space on which the BRST charge $Q'_B$ acts on  
that part of the enlarged space ${\cal H}'_{p.s.}$  
(the pure spinor space which also contains the new fields $\b_m, \k^\a$ and $b$),  
which has non-negative grading (using the same grading  
discussed in the previous sections). This simplifies the comparison with pure-spinor formalism as  
already discussed in the previous section.  
  
\newsec{Zero-Momentum Cohomology} 
 
Another good test of the physical equivalence of our covariant formulation  
with the pure spinor approach is the computation of the zero-momentum cohomology in one  
holomorphic sector.\foot{To restore the complete superstring spectrum, the string field  
for the closed superstring $\Psi_C$ is given by the tensorial product of the two  
sectors $\Psi_C = \Psi_L \otimes \Psi_R$.} We compute the cohomology at zero momentum for all  
ghost numbers. This computation yields all zero-momentum states which describe not  
only the gauge field and its supersymmetric partner, but also, for ghost numbers different from one,  
the target space ghosts and their antifields \bigpicture.  
First, we briefly review the zero momentum cohomology in the pure spinor formulation,  
then we discuss the procedure which extends this result to our formulation and, finally,  
we present the result.  
 
Using pure spinors, the zero-momentum cohomology is described by the string field $\Psi$ 
\eqn\zeroA{ 
\Psi = C \, {\cal U}^{(0)}_0 + a^m \, {\cal U}^{(1)}_{1,m} + \psi^\a \, {\cal U}^{(1)}_{2,\a} +  
+ \psi^*_\a \, {\cal U}^{(2) \a}_{3} + a^*_m \, {\cal U}^{(2) m}_{4} +  
C^* \,  {\cal U}^{(3)}_5\,,  
 } 
where $C,  a_m, \psi^\a$ are the ghost, the gauge field, and the gaugino, while  
$C^*,  a^*_m, \psi^*_\a$ are their antifields. The transversal components of the gauge  
field and the gaugino at $k^m =0$ are the natural extension of the same physical states at  
$k^m \neq 0$, but at $k^m =0$ there are new ``physical'' states which in physical application  
are expected to cancel each other: the longitudinal and timelike components of the Yang-Mills gauge  
and the spacetime Yang-Mills ghost fields.\foot{For the bosonic string the zero-momentum cohomology  
consists of the four states given by $\oint b \, {\cal U} |0, k^m=0 \rangle$ where  ${\cal U} = 1, c^z \p_z x^m,  
c^z \p_z c^z \p_z x^m, c^z \p_z c^z \p^2_z c^z$.}  
The vertices ${\cal U}_i$ generate the  
cohomology $H(Q_B | {\cal H}_{p.s.})$; they are constructed in \BerkovitsRB~and  
are given by  
\eqn\zeroB{ 
{\cal U}^{(0)}_0 = 1\,, ~~~~~~~  
{\cal U}^{(1)}_{1,m} = \l \g^m \t\,, ~~~~~~~ 
{\cal U}^{(1)}_{2,\a} = \l \g^m \t \, (\g_m \t)_\a\,,  
} 
$$ 
{\cal U}^{(2) \a}_{3} =  \l \g^m \t \, \l \g^n \t \, (\g_{mn} \t)^\a \,, ~~~~~~ 
{\cal U}^{(2) m}_{4} =  \l \g^n \t \, \l \g^r \t \, (\t \g_{mnr} \t) \,,  
$$ 
$$ 
{\cal U}^{(3)}_5 =  \l \g^m \t \l \g^n \t \, \l \g^r \t \, (\t \g_{mnr} \t)\,.  
$$ 
The superscripts refer to the ghost number, and all the vertices have  
vanishing conformal spin. An inner product $\langle \Psi, \Psi \rangle$ is  
defined by assuming that the product of the ghost field $C$ should have  
inner product only with its antifield $C^*$, the gauge field $a^m$ with its antifield $a^*_m$ and  
so on.\foot{The definition of an inner product leads to the symplectic BV measure given by  
$\langle \delta \Psi, \delta \Psi \rangle = \int d^{10}x \, \delta\phi^i \wedge \delta\phi^*_i$ where  
$\phi^i$ and the fields and $\phi^*_i$ are the antifields.}  
Therefore, this leads to the conclusion that  
\eqn\zeroC{ 
\langle \, {\cal U}^{(0)}_0, {\cal U}^{(3)}_5 \rangle =  
\langle \,{\cal U}^{(1)}_{1,m}, {\cal U}^{(2) m}_{4} \rangle =  
\langle \, {\cal U}^{(1)}_{2,\a}, {\cal U}^{(2) \a}_{3} \rangle = {\cal N} 
} 
where ${\cal N}$ is a normalization factor. It is easy to check that the vertices \zeroB~indeed  
satisfy the equations \zeroC~and, in particular, by choosing ${\cal N} = 1$ for simplicity, one  
obtains the condition 
\eqn\zeroD{ 
\langle \,  \l \g^m \t \l \g^n \t \, \l \g^r \t \, (\t \g_{mnr} \t) \rangle = 1\,. 
} 
This coincides with Berkovits' prescription for the zero-mode computations in tree level amplitudes \berko and  
it leads to the construction of the measure $\mu(\t,\l)$ for the zero modes at tree level, namely  
\eqn\zeroDA{ 
\langle \,  \l \g^m \t \l \g^n \t \, \l \g^r \t \, (\t \g_{mnr} \t) \rangle  =  
\int \mu(\t,\l)  \Big( \l \g^m \t \l \g^n \t \, \l \g^r \t \, (\t \g_{mnr} \t) \Big) \,, 
} 
$$ 
\mu(\t,\l) = d\Omega_\l \Big( \l^* \g^m {\p \over \p \t }\Big)  
\Big( \l^* \g^n {\p \over \p \t }\Big) \Big( \l^* \g^r {\p \over \p \t }\Big)  
(  {\p \over \p \t }  \g_{mnr} {\p \over \p \t }\Big) \,, 
$$ 
where $d\Omega_\l$ is the Haar measure for the pure spinor coset.  
All vertices in \zeroB~carry a grading (namely the grading of $\l$ is 1 and  
the grading of $\t$ is zero) and since only the ghost $\l^\a$ appears, the  
total grading of each vertex is equal to the ghost number which is positive. Following the  
analysis of the previous sections, given a vertex ${\cal U}^{(n)}_{B,i}$ with ghost number $n$,   
 of the zero momentum cohomology  
$H^{(n)}(Q_B | {\cal H}_{p.s.})$ (the subscript $B$ stands for Berkovits),  
it can be lifted to our cohomology $H^{(n)}(Q | {\cal H}_+)$ such that  
\eqn\zeroE{ 
{\cal U}^{(n)}_i = {\cal U}^{(n)}_{B, i} + \sum^N_{p\geq0} {\cal U}^{(n)}_{(p), i} 
} 
where ${\cal U}^{(n)}_{(p), i}$ is a vertex operator with ghost number $n$ and  
grading number $p$. At zero momentum the charges $Q_2 = 0$ and $Q_3$ have no effect on  
a generic zero momentum vertex ${\cal U}^{(n)}_i$. The latter is a polynomial of $\t$ and of the ghost  
fields $\l^\a,\xi^m,\chi_\a,\omega_m$ and the antighost $b$.  
 
Note that $\{Q_0,  {\cal U}^{(n)}_{B, i} \} = 0$ because ${\cal U}^{(n)}_{B, i}$  
depends only upon $\t$ and $\l$.  
Acting with $Q_B$ on ${\cal U}^{(n)}_{B, i}$ (which coincides with the charge $Q_1$  in \BRSTfinall),   
we obtain that $\{Q_B, {\cal U}^{(n)} \}_{B, i}  = \l\g^m \l M_m(\t)$ where $M_m(\t)$ is a  
polynomial in $\t$. The right hand side is $Q_0$-exact term:   
$\l\g^m \l M_m(\t) = - 2\, \{ Q_0, \xi^m M_m(\t) \}$ since $M_m(\t)$ is $Q_0$ invariant.   
The new vertex operator ${\cal U}^{(n)}_{(n+1), i}  = 2 \, \xi^m M_m(\t) $ has the  
same ghost number as ${\cal U}^{(n)}_{B, i}$, but the grading is increased by one unit. The next step is to  
insert the two vertices in the next equations of the system \specD, namely  
\eqn\zeroF{ 
\{Q_2, {\cal U}^{(n)}_{B, i} \} + \{ Q_B, {\cal U}^{(n)}_{(n+1), i} \} +   
\{ Q_0, {\cal U}^{(n)}_{(n+2), i} \} = 0\,.   
} 
As already pointed out, the first terms are zero, and therefore we have to  
repeat the previous sequence of operations: one has to find  
${\cal U}^{(n)}_{(n+2), i}$ which compensates the $Q_B$ variation of ${\cal U}^{(n)}_{(n+1), i}$.  
At the next level, we have a further equation to satisfy, namely we have 
\eqn\zeroF{ 
\{Q_3, {\cal U}^{(n)}_{B, i} \} + 
\{Q_2, {\cal U}^{(n)}_{(n+1), i} \} +  
\{Q_B, {\cal U}^{(n)}_{(n+2), i} \} +   
\{Q_0, {\cal U}^{(n)}_{(n+3), i} \} = 0\,.   
} 
Again, due to the vanishing of momentum, the action of $Q_2$ and $Q_3$ on  
the vertices ${\cal U}^{(n)}_{B, i}$ and ${\cal U}^{(n)}_{(n+1), i}$ vanishes. Therefore, we can solve  
for ${\cal U}^{(n)}_{(n+3), i}$. At the next level, we have the simplification that by inserting a $b$ term  
we can easily solve the equation. The only limitation comes from the fact that the grading should be  
positive. This means that $n+3 \geq 4$. Finally, we have to take into account that the operation $Q_B$  
removes from ${\cal U}^{(n)}_{(p), i}$  
one fermion $\t$ and replaces it by a ghost $\l$. This means that the new vertex  
${\cal U}^{(n)}_{(p+1), i}$ has one less fermion $\t$ and therefore the sequence of new  
vertices stops when all the $\t$ are removed. This also implies that the highest-grading term  
${\cal U}^{(n)}_{(p), N}$ in the polynomial ${\cal U}^{(n)}_i$ is given by the sum of the  
ghost number $n$ plus the fermion number.  
 
We give two examples. Starting from ${\cal U}^{(1)}_{B, 1,m} = \l \g^m \t$, we have  
$\{Q_B,  \l \g^m \t \} = \l \g^m \l = -  \{Q_0, 2 \xi^m \}$. Here, we have  
${\cal U}^{(1)}_{(2), 1, m} = - 2 \, \xi^m$. Furthemore,  
$\{ Q_B,  -2 \xi^m \} = 0$. This implies that ${\cal U}^{(1)}_{(p), 1, m} = 0$ for all $p\geq 3$. Notice that  
the complete vertex of our cohomology ${\cal U}^{(1)}_{1,m} =  \l \g^m \t  + 2 \, \xi^m$ is not a cohomological  
trivial term at zero momentum. The vertex ${\cal U}^{(1)}_{1,m}$ is coupled to the gauge field $a^m$.  
 
In the same way, starting from ${\cal U}^{(1)}_{2,\a} = \l \g^m \t \, (\g_m \t)_\a$, by using  
the Fierz identities, we have  $\{Q_B, \l \g^m \t \, (\g_m \t)_\a \} = {3\over 2} (\l \g^m \l)(\g_m \t)_\a$.  
This gives the new vertex ${\cal U}^{(1)}_{(2), 2,\a} = - 3 \, \xi^m (\g_m \t)_\a$. Reiterating the procedure,  
we find the new vertex ${\cal U}^{(1)}_{(3), 2,\a} = - 3\, \chi_\a$ and ${\cal U}^{(1)}_{(p), 2,\a} =0$  
for all $p\geq 4$.  
 
The final result is  
\eqn\zeroG{\eqalign{ 
&{\cal U}^{(0)}_0 = 1\,, \cr  
&{\cal U}^{(1)}_{1,m} = \l \g^m \t + 2 \, \xi^m \,, \cr 
&{\cal U}^{(1)}_{2,\a} = \l \g^m \t \, (\g_m \t)_\a - 3 \, \xi^m (\g_m \t)_\a - 3\, \chi_\a\,, \cr 
&{\cal U}^{(2) \a}_{3} = \l \g^m \t \, \l \g^n \t \, (\g_{mn} \t)^\a + 3\, \xi^m (\l \g^n \t) (\g_{mn} \t)^\a  \cr
&~~~~~~~~~ +3\, (\l \g^m \t) (\g_m \chi)^\a + 6\, \xi^m \xi^n (\g_{mn}\t)^\a - 6\, \xi^m (\g_m \chi)^\a\,, \cr 
&{\cal U}^{(2) m}_{4} =  \l \g^n \t \, \l \g^r \t \, (\t \g_{mnr} \t) + 6\, \xi^n \l\g^r \t (\t\g_{rmn} \t)  \cr 
&~~~~~~~~ + 9\, \xi^n \xi^r (\t \g_{rmn} \t) + 6 \, \l \g^r \t (\t \g_r \g_m \chi) + 18 \, \xi^r (\t \g_r \g_m \chi) +  
9 \, \chi \g_m \chi\,, \cr 
&{\cal U}^{(3)}_5 =  \l \g^m \t \l \g^n \t \, \l \g^r \t \, (\t \g_{mnr} \t) + 
8 \, \xi^m (\l \g^n \t) (\l \g^r \t) (\t \g_{mnr} \t)  \cr 
&~~~~~~~~~ + 21\, \xi^m \xi^n (\l \g^r \t) (\t \g_{rmn}\t) 
+ 6\, (\l \g^m \t) (\l \g^r \t) (\t \g_{rm} \chi) + 2 \, \xi^m \xi^n \xi^r (\t \g_{rmn} \t)   \cr 
&~~~~~~~~~ + 18\, (\l \g^m \t) \xi^r (\t \g_{rm} \chi) + 18 \, \xi^m \chi \g_m \chi\,.  
}} 
Note that the first and the last vertex operator are spacetime scalars; this suggests that there are  
no operators with ghost number larger than 3.\foot{Notice that the term  
$\xi^m \xi^n - {1\over 2} \chi \g^{mn} \l$ is BRST invariant, but due to its grading number  
it is also trivial in fact:  
$\xi^m \xi^n - {1\over 2} \chi \g^{mn} \l = \{ Q, b \left(\xi^m \xi^n - {1\over 2} \chi \g^{mn} \l \right) \}$.  
Therefore, it does not belong to the BRST cohomology.} All the operators  
have again vanishing conformal spin.  
 
\newsec{Acknowledgements} 
 
We thank N. Berkovits, W. Siegel, M. Ro\v cek, R. Stora and C. Vafa  
for useful discussions.  P.~A.~G. thanks CERN where the paper has been 
finished.  G. P. thanks C.N. Yang Institute for Theoretical Physics at Stony Brook for  
the hospitality.  This work was partly funded by NSF Grant PHY-0098527 and  
by C.N.R.-Italy.  
 
\vfill 
\footatend\vfill\supereject\immediate\closeout\rfile\writestoppt
\baselineskip=14pt\centerline{{\bf References}}\bigskip{\frenchspacing%
\parindent=20pt\escapechar=` \input refs.tmp\vfill\eject}\nonfrenchspacing 
\bye